\documentclass[final,5p,twocolumn]{elsarticle}
\usepackage{amsmath}
\usepackage{newtxtext,newtxmath}
\usepackage{tikz}
\usepackage{url}
\usepackage{pgfplots}
\usepackage{xcolor,colortbl}
\usepackage{multirow}
\usepackage{xspace}
\usepackage{booktabs}
\usepackage{subcaption}
\usepackage{balance}
\definecolor{Gray}{gray}{0.75}
\def\G{\rowcolor{Gray}}

\newcommand{\MARU}[1]{%
  \ifnum#1>99%
    {\ooalign{\hfil\resizebox{.5\width}{.5\height}{#1}\/\hfil\crcr\lower.2ex\hbox{\mathhexbox20D}}}%
  \else\ifnum#1>9%
    {\ooalign{\hfil\resizebox{.7\width}{.7\height}{#1}\/\hfil\crcr\hbox{\mathhexbox20D}}}%
  \else%
    {\ooalign{\hfil\resizebox{.9\width}{.9\height}{#1}\/\hfil\crcr\raise.1ex\hbox{\mathhexbox20D}}}%
  \fi\fi%
}
\newcommand{\rnum}[1]{\MARU{#1}}

\newcommand{\tuple}[1]{\langle#1\rangle}
\newcommand{\RQ}[1]{\textit{RQ}${}_{#1}$}
\newcommand{\BLT}[1]{\textsf{#1}}
\newcommand{\Mod}[1]{\texttt{#1}}
\newcommand{\CP}[2]{#1: \textit{#2}}
\newcommand{\Config}[1]{$\langle$#1$\rangle$}
\def\Sim{\mathit{Sim}}
\def\Sev{\mathit{Sev}}
\def\nSim{\mathit{nSim}}
\def\nSev{\mathit{nSev}}
\def\nScore{\mathit{nScore}}
\def\nSmell{\mathit{nSmell}}
\def\BLI{\mathit{BLI}}
\def\gBLI{\mathit{gBLI}}
\def\top{\mathrm{top}}
\def\rank{\mathrm{rank}}
\def\Risk{\mathrm{Risk}}

\def\All{\mathrm{all}}
\def\RelativeRisk{\mathrm{RR}}

\def\P{\phantom{0}}
\def\Fst{(1)~}
\def\Snd{(2)~}
\def\Trd{(3)~}

\def\ATFD{\text{ATFD}}
\def\WMC{\text{WMC}}
\def\TCC{\text{TCC}}

\newcommand{\Smell}[1]{\textsf{#1}}
\def\BlobClass{\Smell{Blob Class}\xspace}
\def\DataClass{\Smell{Data Class}\xspace}
\def\DistortedHierarchy{\Smell{Distorted Hierarchy}\xspace}
\def\GodClass{\Smell{God Class}\xspace}
\def\RefusedParentBequest{\Smell{Refused Parent Bequest}\xspace}
\def\SchizophrenicClass{\Smell{Schizophrenic Class}\xspace}
\def\TraditionBreaker{\Smell{Tradition Breaker}\xspace}
\def\BlobOperation{\Smell{Blob Operation}\xspace}
\def\DataClumps{\Smell{Data Clumps}\xspace}
\def\ExternalDuplication{\Smell{External Duplication}\xspace}
\def\FeatureEnvy{\Smell{Feature Envy}\xspace}
\def\IntensiveCoupling{\Smell{Intensive Coupling}\xspace}
\def\InternalDuplication{\Smell{Internal Duplication}\xspace}
\def\MessageChains{\Smell{Message Chains}\xspace}
\def\ShotgunSurgery{\Smell{Shotgun Surgery}\xspace}
\def\SiblingDuplication{\Smell{Sibling Duplication}\xspace}

\newcommand{\tikzv}[1]{\kern-2ex \tikz [remember picture] \node (#1) {};}
\newcommand{\tikzd}[6]{%
  \tikz [overlay,remember picture]
  \draw [decoration={brace,amplitude=2mm,aspect=#6,raise=#4},decorate,thick]
  (#2.south -| rightmark) -- (#1.north -| rightmark)
  node [pos=#6,right=#5,align=right] {#3};
}

\journal{Journal of Systems and Software}
\date{}

\begin{document}
	
\begin{frontmatter}

\title{An Extensive Study on Smell-Aware Bug Localization}

\author{Aoi Takahashi}
\ead{takahashi-a-at@se.cs.titech.ac.jp}

\author{Natthawute Sae-Lim}
\ead{natthawute@se.cs.titech.ac.jp}

\author{Shinpei Hayashi\corref{cor1}}
\ead{hayashi@c.titech.ac.jp}

\author{Motoshi Saeki}
\ead{saeki@se.cs.titech.ac.jp}

\cortext[cor1]{Corresponding author}
\address{School of Computing, Tokyo Institute of Technology, Tokyo, 152--8550 Japan.}

\begin{abstract}
Bug localization is an important aspect of software maintenance because it can locate modules that should be changed to fix a specific bug.
Our previous study showed that the accuracy of the information retrieval~(IR)-based bug localization technique improved when used in combination with code smell information.
Although this technique showed promise, the study showed limited usefulness because of the small number of: 1) projects in the dataset, 2) types of smell information, and 3) baseline bug localization techniques used for assessment.
This paper presents an extension of our previous experiments on Bench4BL, the largest bug localization benchmark dataset available for bug localization.
In addition, we generalized the smell-aware bug localization technique to allow different configurations of smell information, which were combined with various bug localization techniques.
Our results confirmed that our technique can improve the performance of IR-based bug localization techniques for the class level even when large datasets are processed.
Furthermore, because of the optimized configuration of the smell information, our technique can enhance the performance of most state-of-the-art bug localization techniques.
\end{abstract}


\end{frontmatter}


\section{Introduction}\label{s:introduction}
	
Bug localization is the process of identifying the locations of a given bug.
Because it can be a tedious task in large-scale software development projects, many ideas have been proposed to automate this process using software development information.
For instance, we can identify the locations of a bug using the description of bug reports, i.e., information retrieval~(IR)-based~\cite{lukins-ist2010,nguyen-ase2011}, or execution traces, i.e., dynamic analysis~\cite{wong-icsme2014}.
To improve the bug localization accuracy, many hybrid techniques that combine a base technique with additional information have been proposed.
For example, \BLT{BugLocator}~\cite{zhou-icse2012} combined similar bug reports that were fixed in the past with an IR-based technique.
\BLT{BLUiR}~\cite{saha-ase2013} incorporated structural information in addition to using similar bug reports from the past.
\BLT{AmaLgam}~\cite{wang-icpc2014} combined the version history, report similarity, and structural information.

Although these techniques can significantly improve the bug localization accuracy, they can only be used when sufficient additional information is available.
Moreover, most existing techniques do not consider the likelihood of each module containing a bug and treat all modules equally, which may lower the accuracy of bug localization.
To this end, we previously proposed a smell-aware bug localization technique to improve the IR-based bug localization accuracy using code smell information~\cite{takahashi-icpc2018}.
The motivation behind our approach is that modules with code smells have been found to be changed more often and fault-prone~\cite{khomh-emse2012,guerrouj-sqj2017}.
In addition, our technique does not require additional information as code smells can be directly detected from the source code.

Although our technique can significantly enhance the bug localization performance, the previous study still has limitations that need to be addressed.
First, we experimented on only four open-source systems.
This small number of targeted systems means that our results may be difficult to generalize.
Second, we only used one set of configurations for the technique, even though there were many possible options.
Thus, our previously reported result might not be optimal.
Finally, we combined our technique with only one base bug localization technique.
Therefore, it remains unclear whether our technique is applicable to other bug localization techniques.

The study presented in this paper is an extension of our previous study with the objective of overcoming these limitations.
First, we replicated our study on Bench4BL~\cite{lee-issta2018}, which is the largest dataset available for bug localization.
Second, we generalized the smell-aware bug localization technique and conducted studies with different configurations to find the best configurations.
Finally, we combined the smell-aware bug localization technique with different base techniques provided by Bench4BL to assess whether our technique could improve state-of-the-art bug localization techniques.

The main contributions of this paper can be summarized as follows:
\begin{itemize}
	\item We replicate the smell-aware bug localization technique at the class level and show that it is effective even for processing a large-scale dataset.
	\item We generalize the smell-aware bug localization technique to allow different configurations and present the optimal configuration of the technique.
	\item We combine the smell-aware bug localization technique with different base bug localization techniques and show that it can improve their performance.
\end{itemize}

The remainder of this paper is organized as follows.
First, we provide preliminary information about IR-based bug localization and code smells in the next section.
Next, we summarize related work pertaining to empirical studies of bug localization in Section~\ref{s:related_work}.
We describe our smell-aware bug localization technique in Section~\ref{s:technique}.
In Section~\ref{s:experiment}, we provide the details of our study and present the results.
Threats to validity are discussed in Section~\ref{s:threats_to_validity}.
Discussions of this work are presented in Section~\ref{s:discussion}.
Finally, we provide our conclusions in Section~\ref{s:conclusion}.


\section{Background}\label{s:preliminaries}

\subsection{IR-Based Bug Localization and Its Extensions}\label{s:bl}

Bug localization is the process of identifying the location of the source code that should be modified to fix a specific bug.
Bug localization is challenging, especially in large-scale software systems.
Therefore, automated bug localization techniques can help developers save time during such tedious processes.

IR-based bug localization techniques accept a bug report and the source code of a specific version as inputs.
The approaches then determine the similarity between the bug report and source code and generate a ranking of modules based on this similarity.
Developers are expected to use these rankings to help them perform bug-fixing tasks.

In IR-based bug localization, the following steps were conducted to quantify similarity.
\begin{enumerate}
  \item \textbf{Corpus generation.}
  To run the approach, the text of each module needs to be processed.
  The test is regarded as a sequence of tokens.
  In addition, compound words such as \textit{isCommitable} are divided into \emph{is} and \textit{Commitable}\@.
  They are then processed using the standard step as in natural language processing tasks, such as stemming or stop word removal for each module.
  \item \textbf{Indexing.}
  The next step is to perform indexing on the generated corpus.
  Specifically, approaches such as the term frequency-inverse document frequency~(TF-IDF) are applied.
  The approach calculates the importance of each word in each module.
  For example, in the case of using TF-IDF, the importance of each word is calculated by using the frequency of the term.
  \item \textbf{Query construction.}
  To calculate the similarity between the source code and bug report, the bug report is also preprocessed, similar to the corpus generation step.
  \item \textbf{Ranking.}
  The indexed corpus and the bug report are transformed into vectors.
  In the case of the vector space model~(\BLT{VSM}) approach, we can obtain their similarity by calculating the cosine similarity of two vectors.
  The calculated similarity values are then used to rank the modules.
  The higher the rank, the more likely it is to be the location of a bug.
\end{enumerate}

An advantage of IR-based bug localization techniques is that a few types of inputs are required.
As most software development projects currently use an issue tracking system, the bug localization techniques can easily obtain their source code and bug reports.
Therefore, IR-based bug localization can be applied in most situations.
In contrast, a disadvantage is their low accuracy; sufficient accuracy cannot be obtained by merely considering the similarity between the bug report and source code~\cite{zhou-icse2012}.
In addition, this approach depends on the quality of bug reports~\cite{chaparro-icsme2017,kim-tse2013,le-emse2017}.

Because the accuracy of IR-based bug localization is not sufficiently high, many approaches have been proposed to combine it with other information types.
It is noteworthy that the similarity between the bug report and source code is still necessary.
Other types of information are solely an addition.
Types of information that have been applied to bug localization are as follows.
\begin{itemize}
  \item \textbf{Source code size.}
  Zhou et al.~\cite{zhou-icse2012} used the number of lines of source code to represent the size of the source code because bugs are likely to be in the large size source code.
  \item \textbf{Past bug reports.}
  Zhou et al.~\cite{zhou-icse2012} used bug reports similar to current bug reports to support their approach.
  The underlying reason is that similar bug reports are likely to modify the same file.
  \item \textbf{Stack trace.}
  Wong et al.~\cite{wong-icsme2014} used stack trace information in a bug report to capture the order of the executed modules until the program failed.
  Therefore, as we know the modules that were executed, we can use such information to improve bug localization.
  \item \textbf{Change history.}
  Wang et al.~\cite{wang-icpc2014} used the past change history when calculating the probability of buggy modules.
  This information represents the likelihood that a given file will contain a bug in general.
  This information can be calculated by the number of modified files in each bug fixing commit.
\end{itemize}

The advantages and disadvantages of the approaches that combine other information can be the opposite of IR-based approaches.
Because information such as the change history is not always available, the applicability of some approaches is limited.
In contrast, as the approaches use extra information, the accuracy can be improved; for instance, an approach in which the combined size of the source code can significantly improve the IR-based bug localization accuracy~\cite{zhou-icse2012}.

\subsection{Code Smells}\label{s:code-smell}

Code smells are often used as an indicator of a design flaw or problem in the source code~\cite{refactoring}.
Many studies have found that code smells are related to several aspects of software development problems~\cite{yamashita-icsm2012,yamashita-icse2013}.
Thus, it is recommended to remove code smells by performing related refactoring operations to improve the quality of the source code.
Code smells were initially proposed using descriptive language.
Thus, several studies have attempted to implement them formally.

For example, Lanza and Marinescu~\cite{oomip} defined a metric-based strategy for detecting \GodClass as follows:
\begin{align*}
  \text{\GodClass}(m) = {}&(V_\ATFD(m) \ge T_\ATFD) \land{} \\
                          &(V_\WMC(m)  \ge T_\WMC)  \land{} \\
                          &(V_\TCC(m)  \le T_\TCC)
\end{align*}
where $V_\ATFD(m)$, $V_\WMC(m)$, and $V_\TCC(m)$ are the metric values of access to foreign data~(ATFD), weighted method count~(WMC), and tight capsule cohesion~(TCC) of $m$, respectively.
Similarly, $T_\ATFD$, $T_\WMC$, and $T_\TCC$ are the thresholds of ATFD, WMC, and TCC, respectively.
Specifically, ATFD measures the number of foreign attributes that are sued by a class.
Therefore, the higher the ATFD, the more likely the class is to be a \GodClass\@.
Similarly, WMC represents the sum of the complexities of all the methods declared in a class.
Thus, the higher the WMC, the more likely it is that the class is a \GodClass\@.
In contrast, the TCC represents the degree of cohesiveness of a class.
As a result, a class with a lower TCC is more likely to be a \GodClass\@.
These three conditions are then combined as a conjunctive form to determine \GodClass\@.

To measure the strength of a code smell, Marinescu defined \emph{severity} as an integer that measures the number of times the value of a chosen metric exceeds a given threshold~\cite{marinescu-ibmjrd2012}.
For instance, in the case of \GodClass, WMC, ATFD, and TCC were used in the detection approach.
Among these three metrics, ATFD is used to calculate the severity.
In other words, we can calculate the severity by computing the number of times the value of ATFD exceeds its threshold.
Severity values range from 1 to 10.
Note that the metrics used to compute the severity vary according to the smell type because the detection strategy of each smell type uses a different set of metrics.
More information, including additional examples of the severity computation, can be found in the original paper by Marinescu~\cite{marinescu-ibmjrd2012}.


\section{Related Work}\label{s:related_work}

IR-based bug localization is useful for locating source code files that need to be modified to fix a specific bug.
If a bug report and the source code are its inputs, it outputs files or a ranking of files that need to be modified to fix the bug.
IR techniques, e.g., Latent Dirichlet Allocation~(LDA)~\cite{blei2003}, Latent Semantic Indexing~(LSI)~\cite{landauer1998}, or Vector Space Model~(VSM)~\cite{salton1975}, are used for bug localization to calculate the similarity between a given bug report and the source code.
The obtained similarity scores were then utilized to specify the source files from the given bug report.
In addition, techniques of IR-based concept location~\cite{gay-icsm2009,dit-jsep2013} and impact analysis~\cite{gethers-icse2012} follow the same approach.
Among them, Rao and Kak reported \BLT{VSM} to be the best choice among IR techniques for bug localization~\cite{rao-msr2011}.

An advantage of IR-based bug localization techniques is the few types of inputs that need to be prepared because they require only bug reports and source code as their inputs.
However, this could also be regarded as a disadvantage because they depend excessively on the quality of bug reports.
In other words, IR techniques are unable to process low-quality bug reports effectively.
To mitigate this problem, researchers focused directly on the quality of bug descriptions and the tailoring of IR techniques.
Chaparro et al.~\cite{chaparro-icsme2017} proposed a technique to improve IR-based bug localization by reconstructing low-quality bug reports.
Other researchers attempted to rectify the behavior of an IR-based bug localization technique according to the quality of bug reports~\cite{kim-tse2013,le-emse2017}.
Also, Moreno et al.~\cite{moreno-fse2015} proposed a technique named \BLT{QUEST} to automatically configuring the parameters of the used IR approach to improve its accuracy.

Another line of techniques involves the combination of additional information with an IR technique.
Shi et al.~\cite{shi-asc2018} suggested that combining additional information with an IR-based bug localization technique can be beneficial.
Zhou et al.~\cite{zhou-icse2012} defined revised VSM~(\BLT{rVSM}) by considering the scale of the source code in \BLT{VSM}\@.
They proposed \BLT{BugLocator} by combining \BLT{rVSM} with information about similar bug reports from the past.
Wong et al.~\cite{wong-icsme2014} proposed \BLT{BRTracer} by combining it with stack traces.
A stack trace describes the methods that are invoked and the order in which they are invoked until the test fails, which can be obtained by dynamic analysis or reported in bug reports.
Similarly to \BLT{BRTracer}, \BLT{Lobster}~\cite{moreno-icsme2014} also utilized stack trace information in bug reports to improve bug localization.
Tantithamthavorn et al.~\cite{tantithamthavorn-snpd2013} proposed a technique that uses the history of past changes together with \BLT{BugLocator}\@.
In this technique, they used the co-change information in the change history and additionally specified modules that were likely to be changed when a module was changed.
Furthermore, \BLT{BLUiR}~\cite{saha-ase2013} considered the structure of the source code in addition to \BLT{BugLocator}\@.
\BLT{AmaLgam}~\cite{wang-icpc2014} utilizes a bug prediction technique using the version history in addition to \BLT{BLUiR}\@.

Although these hybrid IR-based bug localization techniques are more accurate than basic IR-based bug localization techniques, they are more costly to apply.
For instance, \BLT{AmaLgam} requires the user to collect the change history, which is time-consuming.
In addition, the applicability of techniques using a version history is limited to projects with sufficient history.

\begin{table}[tb]\centering
  \caption{Comparison of existing techniques}\label{t:approach_comparison}
  {\footnotesize\begin{tabular}{lccccccc} \toprule
    Information used  & \rotatebox{90}{\BLT{VSM}} & \rotatebox{90}{\BLT{rVSM}} & \rotatebox{90}{\BLT{BugLocator}} & \rotatebox{90}{\BLT{BRTracer}} & \rotatebox{90}{\BLT{BLUiR}} & \rotatebox{90}{\BLT{AmaLgam}} & \rotatebox{90}{Smell-aware} \\ \midrule
    Similarity        & \checkmark & \checkmark & \checkmark & \checkmark & \checkmark & \checkmark & \\
    Code size         & & \checkmark & \checkmark & \checkmark & \checkmark & \checkmark & \\
    Past bug reports  & & & \checkmark & \checkmark & \checkmark & \checkmark & \\
    Stack trace       & & & & \checkmark & & & \\
    Code structure    & & & & & \checkmark & \checkmark & \\
    Change history    & & & & & & \checkmark & \\
    Smell information & & & & & & & \checkmark \\ \bottomrule
  \end{tabular}}
\end{table}
This study extends our previous study using Bench4BL, which is the largest available dataset for bug localization.
The techniques from Bench4BL that we used in this study are \BLT{BugLocator}, \BLT{BRTracer}, \BLT{BLUiR}, and \BLT{AmaLgam}\@.
In addition, we added \BLT{VSM} as a baseline for bug localization techniques and \BLT{rVSM} as a technique that considers the size of the source code.
The comparison of the information used in each technique is presented in Table~\ref{t:approach_comparison}.
As we can see, most techniques are improved based on another technique by adding more information.
In contrast, our smell-aware technique is dependent on another technique and can be applied on any technique to improve the performance.

As we have already noted in Section~\ref{s:introduction}, the motivation behind our approach was that modules with code smells have been found to be more changed- and fault-prone~\cite{khomh-emse2012,guerrouj-sqj2017}.
Following a similar motivation, there are several attempts to use smell information for bug prediction.
For example, Taba et al.~\cite{taba-icsm2013} have proposed a bug prediction technique that uses a historical metric computed from smelly classes as additional information.
In addition, Palomba et al.~\cite{palomba-tse2017} have succeeded in effectively utilizing the severity degree of code smells in bug prediction.
Their studies suggest the usefulness of smell-based information on identifying buggy portion in source code, which partly justify the use of smell-based information on bug localization.

\section{Smell-Aware Bug Localization}\label{s:technique}

\subsection{Bug Likelihood Index}

The smell-aware bug localization technique aims to improve the accuracy of existing IR-based bug localization.
A problem with IR-based bug localization is that it relies on the textual similarity between a bug report and the source code, i.e., it considers all modules to be equal and does not consider the likelihood of a module containing a bug.
This shortcoming may be responsible for the low accuracy of the technique.
To overcome this problem, our technique uses information about the code smells to represent the likelihood of a module containing a bug.
Specifically, we used the \emph{smell severity}, which indicates the strength of a code smell~\cite{marinescu-ibmjrd2012}, to represent the likelihood.
In addition, as smell information can be directly detected from the source code, it is possible to keep the cost of the technique close to that of IR-based bug localization.
In other words, the user does not need to obtain additional information to use the technique.

\textbf{Example.}
To fix the bug \textsf{CAMEL-9059}\footnote{\url{https://issues.apache.org/jira/browse/CAMEL-9059}}, the method \Mod{createEndpoint} in the class \Mod{JettyHttpComponent} was modified\footnote{\url{https://github.com/apache/camel/commit/a516501}}.
The class had three smells: \RefusedParentBequest, \SchizophrenicClass, and \GodClass\@.
Also, the method had two smells: \FeatureEnvy and \BlobOperation\@.
In particular, \GodClass and \FeatureEnvy were detected as the severest~(severity of 10).
This class module is highly complicated, difficult to comprehend, and difficult to change safely, which might lead to bug-proneness.
Unfortunately, this module was assigned a low rank in IR-based bug localization~(349th in the ranking by \BLT{VSM}) because many other modules were textually more similar to the bug report.
Smell-aware bug localization aims to increase the ranking of these modules by utilizing smell information.

To combine the code smell information with textual similarity, we proposed the \emph{bug likelihood index}~(BLI)\@.
The BLI for each module $m$ can be calculated as follows:
\begin{align*}
  \BLI(m) = (1 - \alpha) \, \nSim(m) + \alpha \, \nSev(m),
\end{align*}
where $\nSim(m)$ is the textual similarity of the bug report and module $m$ based on \BLT{VSM} and $\nSev(m)$ is the sum of the severity of the smells contained in module $m$.
Both values are normalized in the range of $[0, 1]$.
More specifically, they are calculated from the original non-normalized similarity $\Sim(m)$ and severity sum $\Sev(m)$ as follows:
\begin{align*}
  \nSim(m) &= \frac{Sim(m)}{\max_{m' \in M} Sim(m')}, \\
  \nSev(m) &= \frac{Sev(m)}{\max_{m' \in M} Sev(m')}.
\end{align*}
Further, $\alpha~(0 \leq \alpha \leq 1)$ is a parameter representing the weight of $\nSim(m)$ and $\nSev(m)$.

In our previous study, we conducted an experiment using four open-source projects: \textsf{ArgoUML}, \textsf{JabRef}, \textsf{jEdit}, and \textsf{muCommander}\@.
In this study, TraceLab~\cite{bogdan-icpc2012} and inFusion~\cite{marinescu-ibmjrd2012} were used as a \BLT{VSM} implementation and a smell detector, respectively.
In the \BLT{VSM} implementation of TraceLab, TF-IDF was used as its weighting scheme, and it was equipped with a standard IR preprocess.
We applied the IR-based bug localization technique with \BLT{VSM} and our smell-aware bug localization to the targets and compared the accuracy using the mean average precision~(MAP)~\cite{map}.
As a result, our technique could improve IR-based bug localization by 36\%, 34\%, 24\%, and 28\% in relative comparison for \textsf{ArgoUML}, \textsf{JabRef}, \textsf{jEdit}, and \textsf{muCommander}, respectively.

\subsection{Generalized Bug Likelihood Index}

Although the result using BLI was promising, the following limitations must be noted.
\begin{itemize}
	\item \textbf{Generalization.}
	As the goal of our previous study was to obtain a preliminary result to determine whether code smells have the potential to improve bug localization, we conducted our study on a small dataset consisting of four projects.
	Although this dataset is often used in related studies reported in the literature, it is difficult to generalize the applicability of our technique because of the small number of projects.
	\item \textbf{Optimal configuration.}
	The generalization of smell-awareness requires many parameters to be specified.
	For example, we could specify the granularity of code smells, e.g., class or method level.
	We could also specify the aggregator to use when a module has more than one code smell, e.g., summation or obtaining the maximum.
	Finally, as many types of code smells exist, we could choose whether to include all types or only specific types of code smells.
	However, in our previous work, we conducted a study with only one configuration, which means that the reported results might not be optimal.
	\item \textbf{Combination of different base techniques.}
	Smell-aware bug localization was designed to be employed in combination with a base bug localization technique to improve its performance.
	However, in our previous study, \BLT{VSM} was the only base technique to be studied.
	Although \BLT{VSM} is a representative technique for IR-based bug localization because of its simplicity, many other high-performance bug localization techniques have been proposed~\cite{wong-icsme2014,zhou-icse2012,saha-ase2013,wang-icpc2014}.
	Therefore, it remained unclear whether our technique can be used to improve other base techniques.
\end{itemize}

Therefore, to overcome the limitations mentioned above, we generalized the technique to examine different configurations of code smells and defined \emph{generalized bug localization index}~(gBLI).
The gBLI of module $m$ can be calculated as:
\[ \gBLI^{t,c}(m) = (1 - \alpha) \, \nScore^t(m) + \alpha \, \nSmell^c(m) \]
where $\nScore^t(m)$ is the normalized output score of the bug localization technique, $t$.
Furthermore, $\nSmell^c(m)$ is a normalized value based on the code smell configuration $c$, which includes three parameters: granularity~($g$), aggregator~($a$), and type selector~($s$).
The normalization process of $\nScore(m)$ and $\nSmell^c(m)$ is the same as that of $\nSim(m)$ and $\nSev(m)$.
The $\nScore^t(m)$ and $\nSmell^c(m)$ are generalizations of $\nSim(m)$ and $\nSev(m)$, respectively.
The details of the code smell configuration will be explained in the next subsection.

\subsection{Code Smell Configuration}\label{sss:configuration}

A code smell configuration includes three parameters: granularity~($g$), aggregator~($a$), and type selector~($s$), which are explained in the following paragraphs.

\textbf{Granularity~($g$).}
Code smells are often defined on the basis of the granularity of the modules, e.g., the class or method levels.
In our previous study, we used only class-level code smells because we focused on a bug localization technique that outputs class-level results.
In other words, we kept the granularity of the code smells the same as that of the bug localization result.
However, in addition to the class-level code smells, adding method-level smells may improve the performance because these smells add more information to the modules.
In addition, we are likely to obtain a larger number of modules with code smells by considering the method-level code smells.
Nevertheless, this information may add to the noise of the technique and decrease its performance.
Therefore, it might be useful to clarify the effect of using different code smell granularities when applying the technique.

Therefore, in addition to class-level code smells, the use of method-level code smells may improve the performance.
The granularity~($g$) can be set as follows:
\begin{itemize}
  \item \CP{$g_1$}{class level},
  \item \CP{$g_2$}{method level}, and
  \item \CP{$g_3$}{both class and method levels}.
\end{itemize}

\textbf{Aggregator~($a$).}
When detecting code smells, the possibility of more than one code smell being detected in a single module is high.
Therefore, we need an aggregator to combine the information of each code smell to represent the value of the module.
For example, in our previous study, we used summation to combine the severity of each smell in a module.
As another example, Palomba et al.~\cite{palomba-tse2017} used the maximum smell severity value to extract the bug-proneness of a module.
Thus, multiple ways are available in which to combine the smell information.
A comparison of different aggregators is necessary to determine which one performs the best.

We considered the sum and maximum of the severity as aggregators because it was used in the literature.
In addition, we considered using the smell existence, that is, 1 if a module contains at least one smell and 0 otherwise.
We also considered the number of smells in a module.
Furthermore, the average and the median of the severity of all the smells are considered as representatives of the severity degree of the target module.
Finally, considering a situation in which the severity or number of smells is biased depending on the smell types, nested aggregators using the average or median after aggregating by the maximum severity or number of smells are also added.
These aggregators are considered to confirm whether the use of the smell severity yields improved performance.
To summarize, the aggregator~($a$) can be set as follows:
\begin{itemize}
  \item \CP{$a_1$}{sum of severity},
  \item \CP{$a_2$}{maximum severity},
  \item \CP{$a_3$}{existence of smells},
  \item \CP{$a_4$}{number of smells},
  \item \CP{$a_5$}{average of severity},
  \item \CP{$a_6$}{median of severity},
  \item \CP{$a_7$}{average of the maximum severity in each smell type},
  \item \CP{$a_8$}{median of the maximum severity in each smell type},
  \item \CP{$a_9$}{average number of smells in each smell type}, and
  \item \CP{$a_{10}$}{median number of smells in each smell type}.
\end{itemize}

\textbf{Type Selector~($s$).}
Most detectors can detect different types of code smells; for example, inFusion can detect 16 types of code smells.
Our previous study entailed the detection of all types of code smells.
Nevertheless, different types of code smells may affect bug-proneness differently.
Therefore, employing a technique with different types of code smells might yield different results.

Our goal here is to compare the performance when all types of code smells are used with the performance when only certain types of code smells that are more likely to be related to bug proneness are used.
We prepare five setting levels of smell type selection:
\begin{itemize}
  \item \CP{$s_1$}{all smell types},
  \item \CP{$s_2$}{rare selected smell types},
  \item \CP{$s_3$}{medium rare selected smell types},
  \item \CP{$s_4$}{medium selected smell types}, and
  \item \CP{$s_5$}{well selected smell types},
\end{itemize}
which vary on their different inclusiveness of smell types.
The concrete selection of each selector is open at this stage because they will be determined according to an empirical study; see Sections~\ref{ss:rq2} and \ref{ss:rq3}.

\textbf{Specialization to BLI.}
Because the previous technique for class modules used \BLT{VSM} as its base bug localization technique and the $a_1$: sum of the severity of $g_1$: class-level smells of $s_1$: all smell types as its configuration of code smells, gBLI with this configuration can express our original BLI:
\[ \gBLI^{\mathsf{VSM}, \tuple{g_1, a_1, s_1}}(m) \equiv \BLI(m). \]


\section{Empirical Study}\label{s:experiment}

\subsection{Research Questions}

This study focuses on the following four research questions~(RQs).
\begin{itemize}
	\item \textit{\RQ{1}: Does smell-aware bug localization improve IR-based bug localization using VSM even for a large-scale dataset?}
	\item \textit{\RQ{2}: What is the relationship between the performance improvement of the smell-aware bug localization and bug proneness?}
	\item \textit{\RQ{3}: What are the best configurations for smell-aware bug localization as an extension of VSM?}
  \item \textit{\RQ{4}: Is the performance of smell-aware bug localization superior to that of state-of-the-art bug localization techniques?}
\end{itemize}

Details of the motivation for each respective RQ are provided later.

\subsection{Approach Overview}

\begin{figure}[tb]\centering
	\includegraphics[width=1\linewidth]{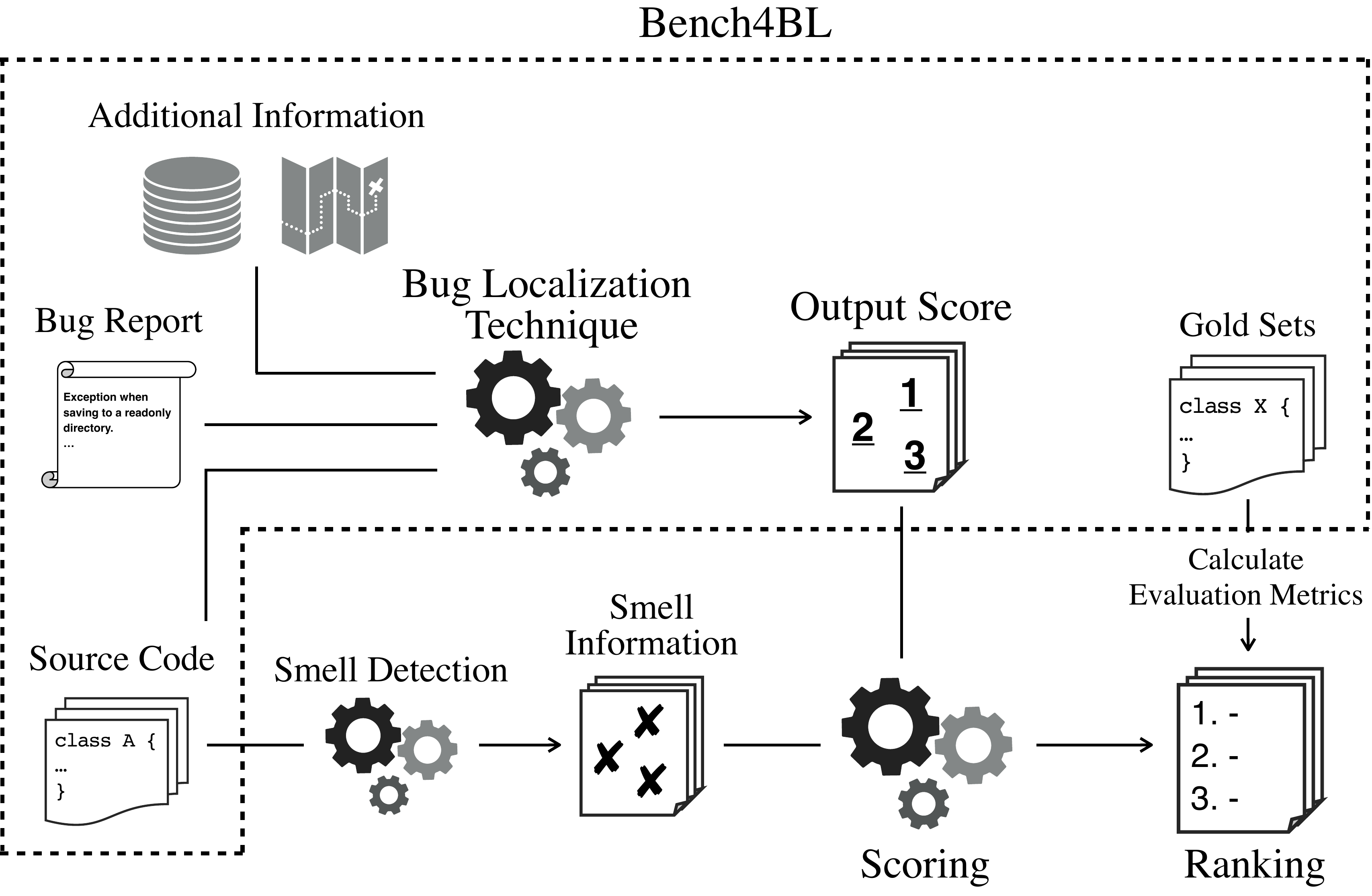}
	\caption{Overview of this study.}\label{f:overview}
\end{figure}
An overview of this study, including the process of Bench4BL and smell detection, is shown schematically in Fig.~\ref{f:overview}.
The nodes within the dotted enclosure are provided by Bench4BL.
First, we executed the bug localization technique using the source code, bug reports, and additional information as inputs.
Next, we detected the code smells from the source code using the aforementioned tool.
Then, we calculated the score of each module from the smell information and output the score obtained by the bug localization technique.
The scores are then used to generate the output ranking.
Finally, we calculated the accuracy of the ranking based on the gold set included in Bench4BL.

\subsection{Data Collection}

\subsubsection{Bench4BL}

In this study, we used Bench4BL, which is the largest benchmark for bug localization~\cite{lee-issta2018}.
The dataset contains the source code, bug reports, and lists of source files that were modified to fix the bug reports, i.e., gold sets, across 46 projects and their versions.
Bench4BL also includes the implementation of state-of-the-art bug localization techniques such as \BLT{BugLocator}~\cite{zhou-icse2012}, \BLT{BRTracer}~\cite{wong-icsme2014}, \BLT{BLUiR}~\cite{saha-ase2013}, \BLT{AmaLgam}~\cite{wang-icpc2014}, \BLT{BLIA}~\cite{youm-ist2017}, and \BLT{Locus}~\cite{wen-ase2016}.
We decided to use Bench4BL because it is suitable for conducting a large-scale empirical study.
Furthermore, it would enable us to combine our smell-aware bug localization with the base techniques implemented in Bench4BL.

We regarded each project version in Bench4BL as a specific system.
In a typical bug localization context, we see a system as a pair of 1) a set of bug reports that define bugs and 2) a source code snapshot to be used to locate the bugs.
When applying this approach to Bench4BL, project versions are most suitable for systems because bug reports are associated with a project version in Bench4BL\@.
This decision also means that we regard different versions of the same project as different systems.

\begin{table*}[tb]\centering
  \caption{Code smells detected by inFusion}\label{t:smell}
  {\footnotesize\begin{tabular}{cll} \toprule
    Granularity & Types & Description \\ \midrule
    \multirow{7}{*}{Class level}
    & \BlobClass~\cite{refactoring,oomip,antipatterns}               & A class that is very large and complex \\
    & \DataClass~\cite{refactoring,oomip,riel-oobook}                & A class with no functionality, only data \\
    & \DistortedHierarchy~\cite{riel-oobook}                         & A class with very narrow and deep inheritance hierarchies \\
    & \GodClass~\cite{refactoring,oomip,riel-oobook}                 & A class that handles data from other classes \\
    & \RefusedParentBequest~\cite{oomip,riel-oobook,martin-agile}    & A class that rarely uses members that inherit from its base class \\
    & \SchizophrenicClass~\cite{riel-oobook,martin-agile}            & A class representing multiple concepts \\
    & \TraditionBreaker~\cite{oomip,riel-oobook}                     & A class that violates the conventions defined by its base class \\ \midrule
    \multirow{9}{*}{Method level}
    & \BlobOperation~\cite{refactoring,oomip,antipatterns}           & A method that is large and complex \\
    & \DataClumps~\cite{refactoring}                                 & A method in which several data values appear as a group \\
    & \ExternalDuplication~\cite{refactoring,antipatterns,hunt-book} & A method containing duplicate code with unrelated classes \\
    & \FeatureEnvy~\cite{refactoring,oomip,riel-oobook}              & A method that is more relevant to the data of other classes than those of its own class \\
    & \IntensiveCoupling~\cite{refactoring,oomip,riel-oobook}        & A method that is more associative with many other methods \\
    & \InternalDuplication~\cite{refactoring,antipatterns,hunt-book} & A method with duplicate code in its own class \\
    & \MessageChains~\cite{refactoring,oomip,antipatterns}           & A method that results in a chain of many method calls \\
    & \ShotgunSurgery~\cite{refactoring,oomip}                       & A method that propagates changes to many other methods when it is changed \\
    & \SiblingDuplication~\cite{refactoring,antipatterns,hunt-book}  & A method with duplicate codes between sibling classes \\ \bottomrule
  \end{tabular}}
\end{table*}

\subsubsection{Bug Localization Techniques}\label{ss:bl-techniques}

Of the six bug localization techniques~(\BLT{BugLocator}, \BLT{BRTracer}, \BLT{BLUiR}, \BLT{AmaLgam}, \BLT{BLIA}, and \BLT{Locus}) provided by Bench4BL, we excluded \BLT{BLIA} and \BLT{Locus} and used the left four techniques for two reasons.
First, these two implementations often behave non-deterministically, outputting different results from the same input.
This behavior was not suitable for our study.
Second, these two implementations had more invalid outputs than the other four implementations.
Following the bug report selection and invalidity criteria shown in Section~\ref{sss:selection}, we could collect 6,936 bug reports that met our criteria from the \BLT{VSM} results.
If we used the four techniques as mentioned above, the number of bug reports decreased to 6,931, which means that only five bug reports were excluded.
However, if we added \BLT{BLIA} into these four, 199 reports were additionally excluded, and 6,732 bug reports remained.
If we added \BLT{Locus} in addition to \BLT{BLIA}, 752 reports were additionally excluded, and 5,980 remained.
Since our goal was to reproducibly confirm the effectiveness of the smell-aware approach on the improved bug localization techniques, not necessarily on all bug localization techniques, we excluded \BLT{BLIA} and \BLT{Locus} to include more bug reports in the experiment.
A similar buggy behavior of the \BLT{Locus} implementation in Bench4BL was also reported by Chaparro et al.~\cite{chaparro-emse2019}.
However, note that the second reason does not directly imply that these implementations are broken.
Some results are valid in terms of bug localization results but inappropriate for our study; see Section~\ref{sss:selection} for the details.

In addition to the four bug localization techniques provided by Bench4BL, we added two techniques: \BLT{VSM} and \BLT{rVSM}.
Both of these techniques are components used in \BLT{BugLocator}, and we used them to ensure that the implementations were compliant with the Bench4BL framework.
\BLT{VSM} is the basis of IR-based bug localization and computes the textual similarity between the bug report and the source code.
We added this technique because it has the lowest cost among bug localization techniques.
We also added \BLT{rVSM}, an extension of \BLT{VSM}, because of our intention to confirm whether smell-aware bug localization is effective only because it considers the size of the source code.
Although we mentioned that code smell is effective for bug localization, the underlying reason may be the size of the source code because certain information from smells represents information about the size of the source code.
Therefore, we compare the smell-aware technique with \BLT{rVSM}, which considers the size of the source code in \RQ{4}.

\begin{figure}[tb]\centering
	\includegraphics[width=0.85\linewidth]{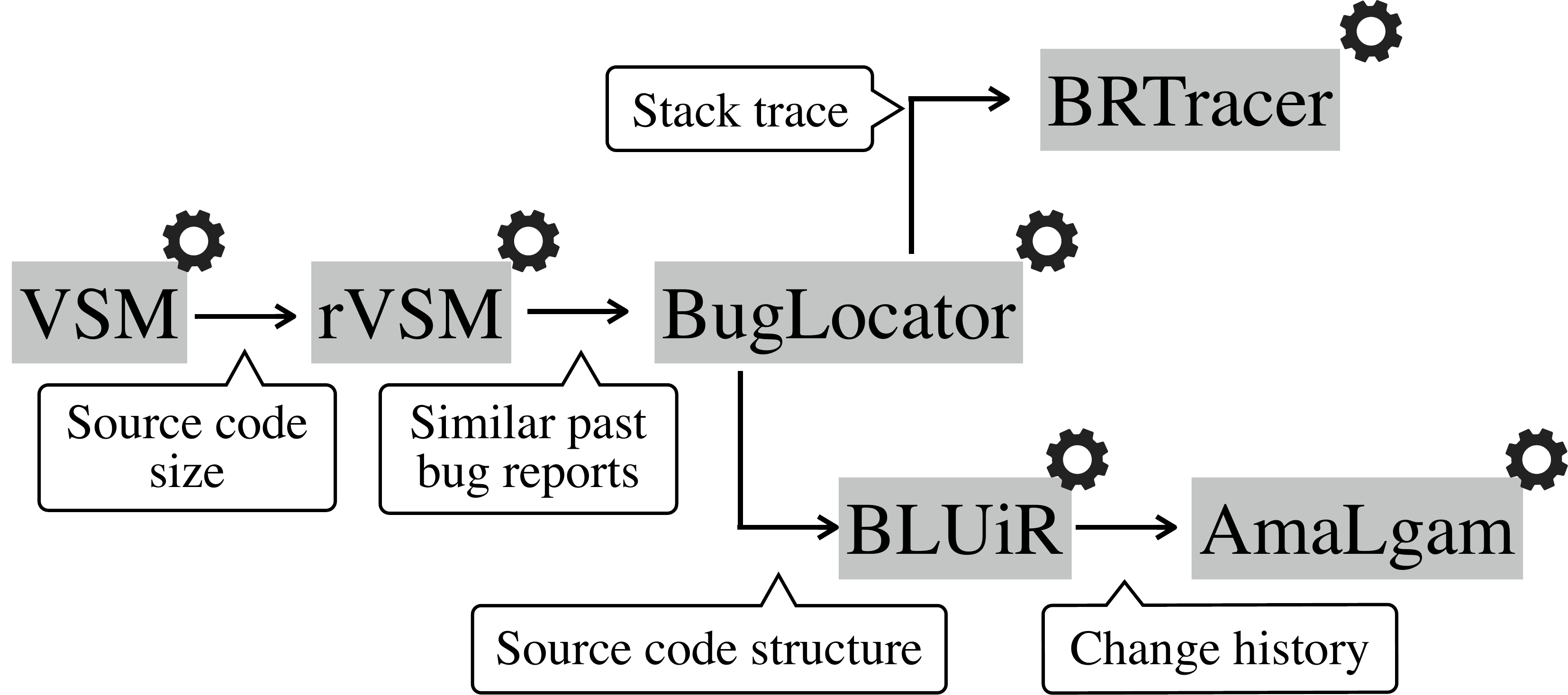}
	\caption{Bug localization techniques used in this study.}\label{f:bl_technique}
\end{figure}
To summarize, we selected the six bug localization techniques shown in Fig.~\ref{f:bl_technique}.
Because all the techniques are based on \BLT{VSM}, we can compare the effects of the additional information contributed by each technique.
\begin{itemize}
	\item \BLT{VSM}~\cite{zhou-icse2012}: The vector space model technique used in \BLT{BugLocator}.
	\item \BLT{rVSM}~\cite{zhou-icse2012}: Extension of \BLT{VSM} that considers the size of the source code.
	\item \BLT{BugLocator}~\cite{zhou-icse2012}: A technique that combines \BLT{rVSM} with previous similar bug reports.
	\item \BLT{BRTracer}~\cite{wong-icsme2014}: A technique that combines \BLT{BugLocator} with extracted stack traces.
	\item \BLT{BLUiR}~\cite{saha-ase2013}: A technique that combines \BLT{BugLocator} with structural information obtained from the source code.
	\item \BLT{AmaLgam}~\cite{wang-icpc2014}: A technique that combines \BLT{BLUiR} with historical information.
\end{itemize}

\subsubsection{Code Smell Detection}

In this study, code smells were detected from the source code of each version in Bench4BL\@.
We used inFusion~\cite{marinescu-ibmjrd2012}\footnote{Because the production company has been closed, it is no longer available. The detected smell instances used in this study are enclosed in the online appendix~\cite{appendix}.}, a powerful commercial code smell detector.
The inFusion is an extended version of inCode~\cite{ganea-scp2017}, which is a successor to iPlasma~\cite{marinescu-icsm2005}.
We selected inFusion for several reasons:
\begin{itemize}
  \item the detected smell instances are associated with a severity score,
  \item it can be assembled in an automated manner without requiring the collection of dependent libraries, compilation of source code, and configuration on an IDE, which helps to simplify the workflow of our experiment with less manual effort,
  \item it follows well-known metric-based smell detection strategies~\cite{oomip}, which are explained in Section~\ref{s:code-smell}, and
  \item it can detect 16 types of code smells at both the class and the method level, which suits our need considering that we aim to compare the effect of different types of code smells in our approach.
\end{itemize}
We regarded the first two reasons as mandatory requirements to conduct our study.
Although other code smell detectors such as cASpER~\cite{stefano-avi2020}, DECOR~\cite{moha-tse2010}, JCodeOdor~\cite{fontana-mtd2015}, and JDeodorant~\cite{tsantalis-csmr2008}, have been proposed to date, they have not met these requirements.

The class-level and method-level smells detected by inFusion are summarized in Table~\ref{t:smell}.
We used the detection result of both of these types of smells without manual validation.

\subsubsection{Data Selection}\label{sss:selection}

We excluded data with any inconsistency from the dataset.
For example, output rankings may be invalid, e.g., those including the similarity score value of Not-a-Number~(NaN) or those with no gold module occurrences.
Although the rankings with no gold module occurrences are valid as bug localization results, we regarded them as invalid because they are not useful for our study in terms of confirming the improvements of the smell-aware bug localization approach.
We excluded any bug reports where the output ranking of at least one bug localization technique was invalid.
In addition, we excluded bug reports of the versions for which inFusion could not detect any code smells because of our intention to confirm the effectiveness of the smell-aware technique.
Finally, we excluded versions with fewer than five bug reports to ensure that each version involved a certain minimum number of bug reports.
This approach was necessary to mitigate the threat of over-optimizing the $\alpha$ values in the case of a small number of bug reports; see Sections~\ref{sss:deciding-alpha} and \ref{ss:internal-validity} for the details.
As a result, we excluded 2,528 of the 9,459 bug reports and used 6,931 bug reports over 309 versions and 35 projects.
Table~\ref{t:projects_information} provides information about each of these projects.
The columns in this table contain the name of the project, the number of versions, the total number of bug reports, and the average number of source files.
\begin{table}[tb]\centering
  \caption{Projects used}\label{t:projects_information}
  {\footnotesize\begin{tabular}{clrrr} \toprule
    \multirow{2}{*}{Group} & \multirow{2}{*}{Project} & \multirow{2}{*}{\# versions} & \# bug & \# files \\
    & & & reports & (mean) \\ \midrule
    \multirow{11}{*}{Apache}
    & CAMEL       &  39 & 1,390 &  9,576 \\
    & HBASE       &  27 &   682 &  2,203 \\
    & HIVE        &  21 & 1,215 &  2,777 \\
    & CODEC       &   4 &    32 &     75 \\
    & COLLECTIONS &   3 &    70 &    468 \\
    & COMPRESS    &  11 &   103 &    174 \\
    & CRYPTO      &   1 &     6 &     82 \\
    & CSV         &   2 &    12 &     26 \\
    & IO          &   7 &    78 &    156 \\
    & LANG        &  10 &   169 &    241 \\
    & MATH        &  10 &   191 &    781 \\ \midrule
    \multirow{4}{*}{JBoss}
    & ENTESB      &   1 &    12 &    252 \\
    & JBMETA      &   1 &    15 &    784 \\
    & SWARM       &   2 &    35 &    493 \\
    & WFCORE      &   8 &   338 &  3,377 \\ \midrule
    \multirow{22}{*}{Spring}
    & AMQP        &   6 &    50 &    266 \\
    & ANDROID     &   1 &    10 &    179 \\
    & BATCH       &  17 &   355 &  1,180 \\
    & BATCHADM    &   2 &    16 &    215 \\
    & DATACMNS    &  12 &    93 &    352 \\
    & DATAGRAPH   &   2 &    12 &    501 \\
    & DATAJPA     &  11 &    66 &    222 \\
    & DATAMONGO   &  19 &   206 &    418 \\
    & DATAREDIS   &   3 &    21 &    331 \\
    & DATAREST    &  10 &    91 &    271 \\
    & LDAP        &   1 &    42 &    433 \\
    & MOBILE      &   1 &     7 &     62 \\
    & ROO         &  11 &   687 &    681 \\
    & SEC         &  27 &   446 &  1,038 \\
    & SECOAUTH    &   1 &    68 &    409 \\
    & SGF         &   7 &    60 &    274 \\
    & SHDP        &   2 &    28 &    956 \\
    & SHL         &   1 &     8 &     89 \\
    & SPR         &   5 &   111 &  4,162 \\
    & SWF         &   9 &    92 &    668 \\
    & SWS         &  14 &   114 &    756 \\ \midrule
    & Total       & 309 & 6,931 &  2,119 \\ \bottomrule
  \end{tabular}}
\end{table}

\subsection{Evaluation Metrics}\label{ss:eval-metric}

To evaluate the ranking outputs from each bug localization technique, we used the following evaluation metrics:
\begin{itemize}
	\item \textbf{Top \textit{N}.}
	This metric represents the ratio in which at least one gold file is included within the top $N$ of the given ranking.
	Here, \emph{gold files} denote files included in the gold set.
	Given a set of bug reports $B$, the metric can be calculated as follows:
	\[ \mathit{Top}~N = \frac{1}{|B|} \sum_{b \in B} \top_N(b) \]
  where $\top_N(b)$ returns 1 if a gold file is contained in the top $N$ of the ranking obtained from bug report $b$, and 0 otherwise.
  By definition, we can calculate the actual number of bug reports that succeeded in having at least one gold module within the top $N$ of the given ranking by multiplying the Top $N$ value by the total number of bug reports.
	In this study, we used Top 1, Top 5, and Top 10.
	
	\item \textbf{Mean Reciprocal Rank~(MRR).}
	MRR~\cite{mrr} is the mean of the multiplicative inverse of the rank of the first gold file in the given ranking.
	Given a bug report $b$, its reciprocal rank~(RR) can be calculated as follows:
	\[ \mathit{RR}(b) = \frac{1}{\rank(b)} \] 
	where $\rank(b)$ is the ranking of the highest gold file in the ranking obtained from bug report $b$.
	Given a set of bug reports $B$, the MRR is calculated as the average of the RR of each bug report in $B$:
	\[ \mathit{MRR} = \frac{1}{|B|} \sum_{b \in B} \mathit{RR}(b). \]
	
	\item \textbf{Mean Average Precision~(MAP).}
	MAP~\cite{map} considers all the gold files, whereas Top $N$ and MRR consider only the top gold files.
	Assuming that the number of files in the output ranking is $N$, the average precision~(AP) of a bug report $b$ can be calculated as follows:
	\[ \mathit{AP}(b) = \sum_{i=1}^N \frac{ \mathit{precision}_b(i) \times \mathit{gold}_b(i) }{ \text{number of gold files} }. \]
	Here, $i$ denotes the rank of a file and $\mathit{precision}_b(i)$ denotes the ratio of gold files in the files ranked at or higher than the $i$-th rank.
	$\mathit{gold}_b(i)$ returns 1 if the file ranked at the $i$-th position is in the gold set and 0 otherwise.
	MAP is the average of AP of all the bug reports $B$:
	\[ \mathit{MAP} = \frac{1}{|B|} \sum_{b \in B} \mathit{AP}(b). \]
\end{itemize}

Moreover, we used the Wilcoxon signed-rank test~\cite{wilcoxon} for statistical testing.
Because the values of all the evaluation metrics are computed as the average of the values for each bug report, we used the set of values for each bug report for all the statistical tests in this study.
For example, when testing the statistical significance of the difference between the two techniques in terms of MAP aspect, we compared two sets of AP values, which were used to compute MAP.
In addition, when reporting the statistical significance, we also reported Cliff's delta~($d$) as a measure of the magnitude of the improvement.
The Cliff's delta is interpreted based on the threshold by Romano et al.~\cite{romano-fair2006}:
  \textit{negligible} for $|d| < 0.147$,
  \textit{small}      for $0.147 \leq |d| < 0.33$,
  \textit{medium}     for $0.33 \leq |d| < 0.474$, and
  \textit{large}      for $0.474 \leq |d|$.

\subsection{\RQ{1}: Does smell-aware bug localization improve IR-based bug localization using VSM even for a large-scale dataset?}\label{ss:rq1}

\subsubsection{Motivation}

Although our previous study showed that smell-aware bug localization can significantly improve the accuracy of IR-based bug localization, the study was conducted with only four projects.
In addition, the number of bug reports included in the dataset was only 277, which made it difficult to generalize the results.
Therefore, the goal of this RQ is to verify whether the smell-aware bug localization technique can be used to improve the performance even for a large-scale dataset.
This is intended as a sanity check to confirm whether the same setting involving the use of the original smell-aware bug localization would be applicable to systems in Bench4BL prior to making new attempts.

\subsubsection{Study Design}\label{sss:deciding-alpha}

This study was designed to replicate and extend the original study to all projects in Bench4BL.
The smell-aware bug localization technique can be employed by setting the smell granularity to either the class or method level.
We used the smell granularity at the class level because the granularity of modules obtained in bug localization is at the file level in Bench4BL.
In addition, we used a Java file as a proxy for the class and excluded all inner classes.
The $\alpha$ value was set to a value such that each evaluation metric is maximized in each system, which ensures that the study is conducted under the same conditions as in the previous experiment.
We produced the rankings for each system by calculating the BLIs from the bug reports in the system using all the $\alpha$ possibilities, which range from 0 to 1 in increments of 0.01.
We then evaluated the set of rankings for each $\alpha$ by employing the evaluation metric that was used.
Finally, the $\alpha$ value that maximized the evaluation score was used as the parameter for the pair of the evaluation metric and the system.

For each bug report, all files in the bug localization result were sorted in descending order of BLI, and the accuracy of IR-based bug localization and the smell-aware bug localization technique were compared according to the gold set in Bench4BL.

\begin{table*}[t]\centering
  \caption{Application example to \textsf{HBASE-1795}.}\label{t:app-example}
  \begin{subtable}[t]{.52\textwidth}\centering
    \caption{\BLT{VSM}~($\alpha = 0$)}
    {\footnotesize\begin{tabular}{rlccc} \hline
      Rank & Module & $\nScore$ & $\nSmell$ & $\BLI$ \\ \hline
       1 & \Mod{TestTHLog}                       & 1.000 & 0.000 & 1.000 \\
       2 & \Mod{THLogRecoveryManager}            & 0.949 & 0.000 & 0.949 \\
       3 & \Mod{HLog}                            & 0.792 & 0.231 & 0.792 \\
       4 & \Mod{TransactionalRegion}             & 0.784 & 0.000 & 0.784 \\
       5 & \Mod{TestStoreScanner}                & 0.659 & 0.000 & 0.659 \\
       6 & \Mod{HRegion}                         & 0.642 & 0.769 & 0.642 \\
       7 & \Mod{BatchMutation}                   & 0.637 & 0.000 & 0.637 \\
       8 & \Mod{TestTHLogRecovery}               & 0.619 & 0.000 & 0.619 \\
       9 & \Mod{TestMinorCompactingStoreScanner} & 0.600 & 0.000 & 0.600 \\
    \G10 & \Mod{Store}                           & 0.582 & 1.000 & 0.582 \\ \hline
	\end{tabular}}
  \end{subtable}
  \begin{subtable}[t]{.45\textwidth}\centering
    \caption{Smell-aware approach~($\alpha = 0.31$)}
    {\footnotesize\begin{tabular}{rlccc} \hline
      Rank & Module & $\nScore$ & $\nSmell$ & $\BLI$ \\ \hline
     \G1 & \Mod{Store}                & 0.582 & 1.000 & 0.711 \\
       2 & \Mod{TestTHLog}            & 1.000 & 0.000 & 0.690 \\
       3 & \Mod{HRegion}              & 0.642 & 0.769 & 0.681 \\
       4 & \Mod{THLogRecoveryManager} & 0.949 & 0.000 & 0.655 \\
       5 & \Mod{HLog}                 & 0.792 & 0.231 & 0.618 \\
       6 & \Mod{TransactionalRegion}  & 0.784 & 0.000 & 0.541 \\
       7 & \Mod{ServerManager}        & 0.364 & 0.769 & 0.489 \\
       8 & \Mod{IndexedRegion}        & 0.325 & 0.769 & 0.462 \\
       9 & \Mod{TestStoreScanner}     & 0.659 & 0.000 & 0.455 \\
      10 & \Mod{BatchMutation}        & 0.637 & 0.000 & 0.440 \\ \hline
    \end{tabular}}
  \end{subtable}
\end{table*}

\subsubsection{Results}

\begin{figure}[tb]\centering
	\includegraphics[width=0.7\linewidth]{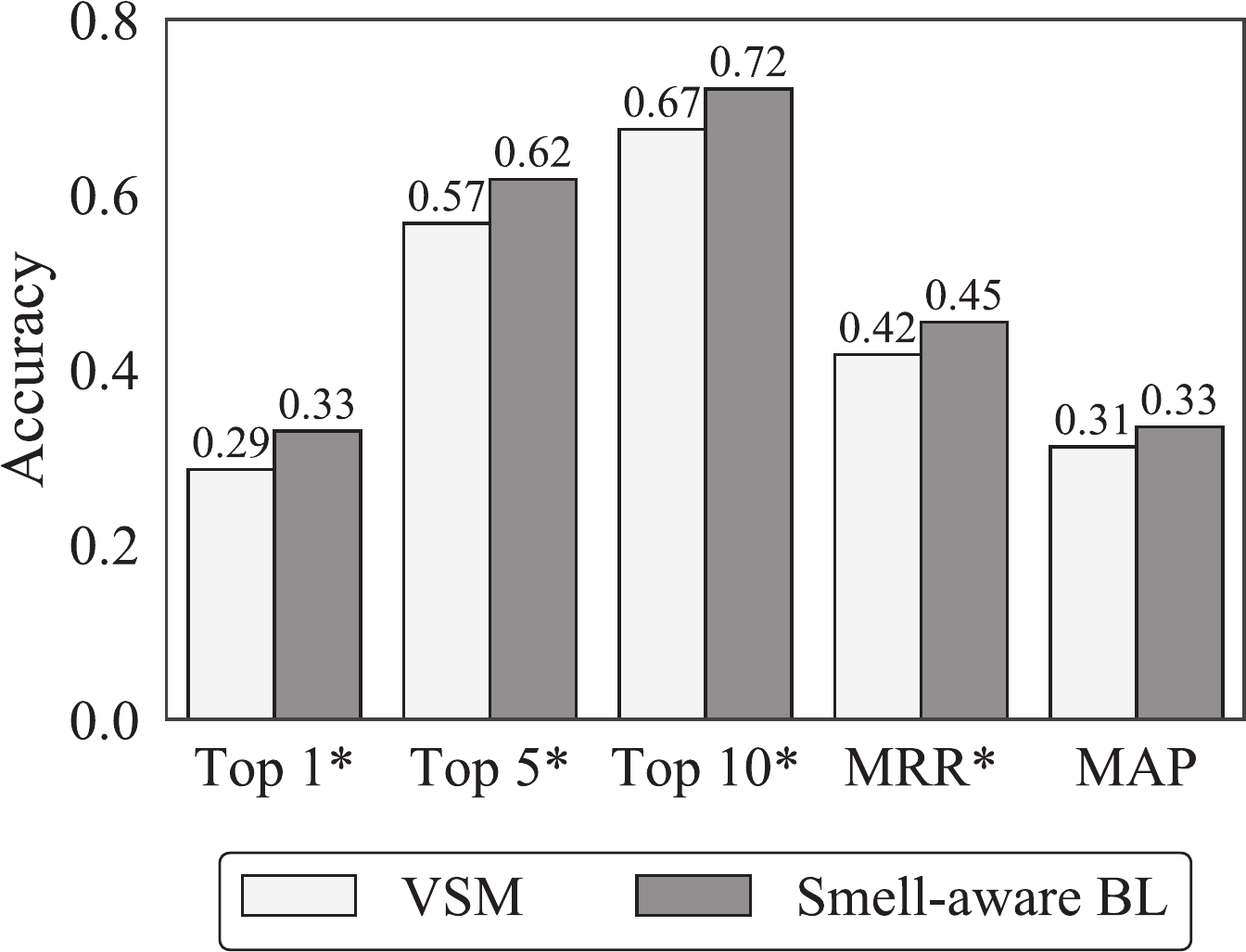}
	\caption{Comparison of the accuracy of smell-aware bug localization and VSM.}\label{f:orig-vsm}
\end{figure}
The results are shown in Fig.~\ref{f:orig-vsm}.
Clearly, the bug localization results for all the evaluation metrics improved.
The increase ratios in terms of the Top 1, Top 5, Top 10, MRR, and MAP metrics were 15.4\%, 8.9\%, 6.9\%, 8.9\%, and 7.3\% in relative comparison~(0.044, 0.050, 0.046, 0.037, and 0.023 in absolute comparison), respectively.
The metrics that are statistically significant~($p < 0.01$) are denoted by an asterisk~(*).
All the differences were statistically significant except for the improvement in MAP: $p=0.13$~(Cliff's delta: 0.044, 0.050, 0.046, 0.051, and 0.042, all \emph{negligible}).
For example, Top 10 increased by 0.046~(from 0.675 to 0.721), which means that the total number of bug reports with gold modules in their top 10-ranked items increased by 321~(from 4,676 to 4,997).

It is noteworthy that MAP in this study increased by a lesser amount than in the previous study.
Specifically, in this study, MAP increased by 7.3\% on average in relative comparison, whereas in the previous study, it was 30.5\%.
The difference may be attributed to several factors.
First, the previous experiment involved all the classes, including the inner classes, whereas this experiment considered only the top-level classes and excluded the inner classes.
Next, the \BLT{VSM} implementation we used differed from that in the previous study.
The \BLT{VSM} implementation in this study is the VSM part of \BLT{BugLocator}, which was optimized for bug localization usage~\cite{zhou-icse2012}.
In addition, the number of projects and bug reports used in this study is much larger than in the previous study; i.e., 4 vs.\ 36 projects and 277 vs.\ 6,943 bug reports were used previously and in this study, respectively.
Consequently, the result of our previous study might be an extreme case, whereas the results of this study reflect a more realistic distribution.

\textbf{Example.}
The application of smell-aware bug localization to \textsf{HBASE-1795}\footnote{\url{https://issues.apache.org/jira/browse/HBASE-1795}} is presented in Table~\ref{t:app-example}.
In Table~\ref{t:app-example}, the top 10 items of the result using \BLT{VSM}~($\alpha = 0$) and that using the smell-aware bug localization technique~($\alpha = 0.31$) are compared.
The gold module, i.e., the class module that was modified to fix this bug, is highlighted in gray.
The gold module \Mod{Store}, which is located at the tenth rank in \BLT{VSM}, has several class-level smells such as \BlobClass and \GodClass\@.
Therefore, \Mod{Store} eventually had the top $\nSmell$ score and was ranked at the top when using smell-aware bug localization.
In addition to \textsf{HBASE-1795}, a total of 24 bug reports tied to \textsf{HBASE 0.20.5} targeted \Mod{Store} to fix, and the score improvement of \Mod{Store} led to improved rankings for many in the system.
Note that in addition to the class-level smells, certain methods of \Mod{Store} have a method-level smell called \BlobOperation, which could also be utilized in smell-aware bug localization.
The use of such method-level smells was studied in \RQ{3}.

\textbf{In summary, the smell-aware bug localization technique at the class level can also improve the accuracy of IR-based bug localization using VSM even for a large dataset.}

\subsection{\RQ{2}: What is the relationship between the performance improvement of the smell-aware bug localization and bug proneness?}\label{ss:rq2}

\begin{table*}[tb]\centering
  \caption{Relative risks of the number of buggy modules included in smelly modules}\label{t:smell_ratio}
  \hspace{5em}
  {\footnotesize\tabcolsep=5pt\begin{tabular}{lrrrrr} \toprule
    \tikzv{rightmark}$t$: Type & \# modules & \# buggy & $\Risk_t$~(\%) & $\Risk^*_t$~(\%) & $\RelativeRisk_t$ \\ \midrule \tikzv{beg}
    \BlobClass              &  1,151 &   246 & 21.373 & 2.385 & (1)  8.960 \\
    {\ShotgunSurgery}*      &    357 &    59 & 16.527 & 2.411 & (2)  6.855 \\
    \GodClass               &  6,108 &   959 & 15.701 & 2.294 & (3)  6.846 \\
    {\BlobOperation}*       &  6,802 & 1,021 & 15.010 & 2.286 & (4)  6.565 \\ \tikzv{s5}
    {\IntensiveCoupling}*   &  2,683 &   290 & 10.819 & 2.384 & (5)  4.534 \\ \arrayrulecolor{black!30}\midrule
    {\DataClumps}*          &  8,928 &   702 &  7.863 & 2.343 & (6)  3.355 \\
    \RefusedParentBequest   &  1,519 &   119 &  7.834 & 2.406 & (7)  3.256 \\
    {\InternalDuplication}* &  3,958 &   306 &  7.731 & 2.386 & (8)  3.240 \\
    {\ExternalDuplication}* &  6,875 &   501 &  7.287 & 2.367 & (9)  3.079 \\ \tikzv{s4} 
    {\FeatureEnvy}*         &  9,614 &   669 &  6.959 & 2.351 & (10) 2.960 \\ \midrule
    {\MessageChains}*       &  2,641 &   176 &  6.664 & 2.401 & (11) 2.775 \\ \tikzv{s3}
    \SchizophrenicClass     &  3,675 &   222 &  6.041 & 2.398 & (12) 2.519 \\ \midrule
    \TraditionBreaker       &  2,159 &    74 &  3.428 & 2.415 & (13) 1.419 \\ \tikzv{s2}
    {\SiblingDuplication}*  &  7,860 &   267 &  3.397 & 2.407 & (14) 1.411 \\ \midrule
    \DataClass              & 16,028 &   361 &  2.252 & 2.423 & (15) 0.930 \\ \tikzv{s1}
    \DistortedHierarchy     &      5 &     0 &  0.000 & 2.419 & (16) 0.000 \\ \arrayrulecolor{black}\midrule
    Total~(all smell types) & 63,953 & 3,668 &  5.735 & 2.060 &      2.785 \\
    All code files          &654,674 &15,834 &  2.419 &    -- &         -- \\ \bottomrule
    \multicolumn{3}{l}{*: method-level}
  \end{tabular}}
  \tikzd{beg}{s1}{{\scriptsize$s_1$}}{6pt}{-25pt}{0.1}%
	\tikzd{beg}{s2}{{\scriptsize$s_2$}}{14pt}{-33pt}{0.1}%
	\tikzd{beg}{s3}{{\scriptsize$s_3$}}{22pt}{-41pt}{0.1}%
	\tikzd{beg}{s4}{{\scriptsize$s_4$}}{30pt}{-49pt}{0.3}%
  \tikzd{beg}{s5}{{\scriptsize$s_5$}}{38pt}{-57pt}{0.5}%
  \hspace{-0.5em}%
  {\footnotesize\tabcolsep=5pt\begin{tabular}{rrr} \toprule
    $\RelativeRisk_t$~(Apache) & $\RelativeRisk_t$~(JBoss) & $\RelativeRisk_t$~(Spring) \\ \midrule
    (1)   11.401 & (2)  \P6.927 & (2)  5.100 \\
    (3)  \P8.465 & (1)   29.963 & (4)  4.589 \\
    (4)  \P8.407 & (3)  \P6.101 & (1)  5.945 \\
    (2)  \P8.539 & (5)  \P3.555 & (3)  4.617 \\
    (5)  \P5.499 & (4)  \P5.199 & (6)  3.536 \\ \arrayrulecolor{black!30}\midrule
    (8)  \P3.921 & (6)  \P2.898 & (9)  2.810 \\
    (9)  \P3.782 & (9)  \P1.965 & (7)  3.401 \\
    (6)  \P4.387 & (7)  \P2.150 & (14) 1.174 \\
    (7)  \P4.142 & (12) \P1.226 & (11) 1.508 \\
    (10) \P3.479 & (10) \P1.732 & (10) 2.542 \\ \midrule
    (11) \P3.365 & (15) \P0.000 & (5)  4.207 \\
    (12) \P3.005 & (13) \P0.981 & (8)  3.134 \\ \midrule
    (14) \P1.670 & (8)  \P2.056 & (12) 1.323 \\
    (13) \P1.672 & (11) \P1.593 & (13) 1.232 \\ \midrule
    (15) \P0.944 & (14) \P0.610 & (15) 0.903 \\
              -- & (15) \P0.000 & (16) 0.000 \\ \arrayrulecolor{black}\midrule
           3.528 &        2.481 &      1.969 \\
              -- &           -- &         -- \\ \bottomrule
    ~
  \end{tabular}}%
\end{table*}

\subsubsection{Motivation}

We showed that smell-aware bug localization could also improve the bug localization accuracy even for the Bench4BL dataset in answering \RQ{1}.
As already explained in Section~\ref{s:introduction}, we consider that the results of smell-aware bug localization are improved because smells are bug prone~\cite{khomh-emse2012,guerrouj-sqj2017}.
To provide more convincing evidence of bug-proneness in the bug localization context in this study and investigate the difference in contributions by the smell types, we examine the extent to which each smell type affects the possibility of identifying buggy modules by bug localization.

\subsubsection{Study Design}

To answer \RQ{2}, we calculate the \emph{relative risk}~\cite{sistrom-radiology2004} of the existence of buggy modules for each smell type.
Let $M_\All$ and $B_\All \subseteq M_\All$ be the sets of all the modules and the \emph{buggy} modules for all the target 309 project versions, where each element is represented as a pair of a module name and the project version to which it belongs.
Here, we regarded a module as buggy if and only if the module was included in the gold set of at least one bug report in the target version in Bench4BL.
We denote the set of modules in which a smell of type $t$ is detected by $M_t \subseteq M_\All$.
The set of buggy modules that contains the smell of $t$ is computed as $B_t = M_t \cap B_\All$.
Then, the risk of smelly modules that are likely to contain bugs to be fixed~($\Risk_t$), that of non-smelly modules~($\Risk^*_t$), and the relative risk of smelly modules~($\RelativeRisk_t$) are, respectively, expressed as follows:
\begin{align*}
  \Risk_t         &= \frac{|B_t|}{|M_t|}, &
  \Risk^*_t       &= \frac{|B_\All \setminus B_t|}{|M_\All \setminus M_t|}, &
  \RelativeRisk_t &= \frac{\Risk_t}{\Risk^*_t}.
\end{align*}

\subsubsection{Results}

The results are presented in Table~\ref{t:smell_ratio}.
The columns show
  the smell type,
  the number of detected smelly modules~($|M_t|$),
  the number of buggy modules in the files detected as smelly~($|B_t|$),
  the buggy risk of the smelly modules~($\Risk_t$) and non-smelly modules~($\Risk^*_t$),
  and the buggy relative risk~($\RelativeRisk_t$) for all systems.
The smell types are sorted in descending order of the obtained relative risks of the smelly modules.
The total number of modules~($|M_\All| = {}$654,674) and buggy modules~($|B_\All| = {}$15,834) are also shown in the bottom row of the table.
The results in the table indicate that, when considering all smell types, 5.735\% of the smelly modules are buggy modules.
The relative risk shows that, in comparison with non-smelly modules, smelly modules are 2.785 times more likely to be buggy.
This suggests that prioritizing smell-containing modules in bug localization can lead to improved accuracy.

The relative risk of smelly modules varied depending on their smell type.
On the one hand, \BlobClass, \ShotgunSurgery, \GodClass, \BlobOperation, and \IntensiveCoupling were the smell types with the top five relative risks.
The risk of being identified as bugs in modules with these smell types is more than four times higher than modules without these smells.
On the other hand, the relative risks for \DataClass and \DistortedHierarchy were less than 1, such that the choices of these smells do not necessarily lead to the identification of modules with a high probability of containing bugs.

The three columns on the right in Table~\ref{t:smell_ratio} present the relative risk of each smell type calculated using only the systems of a specific project group to determine the extent to which the obtained trend is universal.
The numbers in parentheses in the table indicate the rank of an item in the project group to which it belongs.
As we can see from the table, although several small differences exist, the ranking trend for each project group is similar to the global ranking.

\textbf{In summary, in comparison with non-smelly modules, smelly modules are 2.785 times more likely to be buggy.}

\begin{table*}[tb]\centering
  \caption{Accuracy of top 20 configurations}\label{t:conf-perf}
  {\footnotesize\begin{tabular}{llllrrrrrr} \toprule
    Rank     & $g$: Granularity   & $a$: Aggregator             & $s$: Type selector &      Top 1 &      Top 5 &     Top 10 &        MRR &        MAP & \# systems \\ \midrule
   \rnum{0}  & (Ideal)            & (Ideal)                     & (Ideal)            &     0.3864 &     0.6628 &     0.7595 &     0.4960 &     0.3586 &        257 \\ \arrayrulecolor{black!30}\midrule
   \rnum{1}  & \CP{$g_3$}{both levels} & \CP{$a_3$}{existence}            & \CP{$s_5$}{well}        & \Fst0.3515 & \Fst0.6325 &     0.7311 & \Fst0.4707 & \Fst0.3441 &        201 \\
   \rnum{2}  & \CP{$g_3$}{both levels} & \CP{$a_2$}{maximum severity}     & \CP{$s_3$}{medium rare} & \Snd0.3470 &     0.6296 & \Fst0.7348 & \Snd0.4682 & \Snd0.3435 &        229 \\
   \rnum{3}  & \CP{$g_3$}{both levels} & \CP{$a_2$}{maximum severity}     & \CP{$s_4$}{medium}      &     0.3461 & \Trd0.6304 & \Trd0.7337 & \Trd0.4676 & \Trd0.3428 &        220 \\
   \rnum{4}  & \CP{$g_3$}{both levels} & \CP{$a_2$}{maximum severity}     & \CP{$s_1$}{all}         &     0.3464 & \Snd0.6311 & \Snd0.7344 &     0.4675 &     0.3427 &        240 \\
   \rnum{5}  & \CP{$g_3$}{both levels} & \CP{$a_2$}{maximum severity}     & \CP{$s_2$}{rare}        &     0.3451 &     0.6296 &     0.7331 &     0.4671 &     0.3424 &        224 \\
   \rnum{6}  & \CP{$g_3$}{both levels} & \CP{$a_2$}{maximum severity}     & \CP{$s_5$}{well}        &     0.3441 &     0.6263 &     0.7308 &     0.4662 &     0.3419 &        202 \\
   \rnum{7}  & \CP{$g_3$}{both levels} & \CP{$a_3$}{existence}            & \CP{$s_2$}{rare}        & \Trd0.3466 &     0.6301 &     0.7296 &     0.4661 &     0.3411 &        230 \\
   \rnum{8}  & \CP{$g_3$}{both levels} & \CP{$a_3$}{existence}            & \CP{$s_3$}{medium rare} &     0.3460 &     0.6279 &     0.7302 &     0.4653 &     0.3411 &        230 \\
   \rnum{9}  & \CP{$g_3$}{both levels} & \CP{$a_3$}{existence}            & \CP{$s_4$}{medium}      &     0.3453 &     0.6286 &     0.7292 &     0.4653 &     0.3410 &        220 \\
   \rnum{10} & \CP{$g_3$}{both levels} & \CP{$a_9$}{ave. \# of smells}    & \CP{$s_5$}{well}        &     0.3425 &     0.6231 &     0.7266 &     0.4631 &     0.3404 &        200 \\
   \rnum{11} & \CP{$g_3$}{both levels} & \CP{$a_7$}{ave. max severity}    & \CP{$s_5$}{well}        &     0.3411 &     0.6205 &     0.7240 &     0.4622 &     0.3398 &        203 \\
   \rnum{12} & \CP{$g_3$}{both levels} & \CP{$a_8$}{med. max severity}    & \CP{$s_5$}{well}        &     0.3412 &     0.6207 &     0.7234 &     0.4622 &     0.3397 &        203 \\
   \rnum{13} & \CP{$g_1$}{class level} & \CP{$a_3$}{existence}            & \CP{$s_4$}{medium}      &     0.3415 &     0.6226 &     0.7228 &     0.4635 &     0.3393 &        167 \\
   \rnum{14} & \CP{$g_1$}{class level} & \CP{$a_3$}{existence}            & \CP{$s_3$}{medium rare} &     0.3412 &     0.6226 &     0.7236 &     0.4626 &     0.3392 &        185 \\
   \rnum{15} & \CP{$g_1$}{class level} & \CP{$a_{10}$}{med. \# of smells} & \CP{$s_4$}{medium}      &     0.3409 &     0.6221 &     0.7224 &     0.4632 &     0.3391 &        167 \\
   \rnum{16} & \CP{$g_1$}{class level} & \CP{$a_9$}{ave. \# of smells}    & \CP{$s_4$}{medium}      &     0.3409 &     0.6221 &     0.7224 &     0.4632 &     0.3391 &        167 \\
   \rnum{17} & \CP{$g_1$}{class level} & \CP{$a_9$}{ave. \# of smells}    & \CP{$s_3$}{medium rare} &     0.3406 &     0.6223 &     0.7230 &     0.4623 &     0.3391 &        185 \\
   \rnum{18} & \CP{$g_1$}{class level} & \CP{$a_{10}$}{med. \# of smells} & \CP{$s_3$}{medium rare} &     0.3406 &     0.6223 &     0.7230 &     0.4623 &     0.3391 &        185 \\
   \rnum{19} & \CP{$g_1$}{class level} & \CP{$a_3$}{existence}            & \CP{$s_2$}{rare}        &     0.3406 &     0.6227 &     0.7238 &     0.4623 &     0.3389 &        186 \\
   \rnum{20} & \CP{$g_1$}{class level} & \CP{$a_{10}$}{med. \# of smells} & \CP{$s_5$}{well}        &     0.3393 &     0.6205 &     0.7221 &     0.4620 &     0.3389 &        162 \\ \midrule
   \rnum{74} & \CP{$g_1$}{class level} & \CP{$a_1$}{sum of severity}      & \CP{$s_1$}{all}         &     0.3295 &     0.6177 &     0.7210 &     0.4542 &     0.3345 &        204 \\ \arrayrulecolor{black}\bottomrule
 \end{tabular}}
\end{table*}

\subsection{\RQ{3}: What are the best configurations for smell-aware bug localization as an extension of VSM?}\label{ss:rq3}

\subsubsection{Motivation}

As discussed above, when the smell-aware bug localization technique was used in our previous work, we used only one set of configurations, even though many options were available.
For example, we can change the granularity of code smells, the aggregator when combining multiple code smells, and the type of code smells.
Therefore, when answering this RQ, our goal was to explore the configurations that would yield the best performance.
 
\subsubsection{Study Design}

In our previous study, the technique was limited to textual similarity~($\nSim$) and the sum of the severity~($\nSev$) as shown when formulating BLI\@.
To answer \RQ{3}, we utilized the gBLI with three parameters of the code smell configuration $c$ defined in Section~\ref{s:technique}, as a combination of three granularity levels~($g$), ten aggregators~($a$), and five type selectors~($s$).

We instantiated concrete selections of type selectors.
Based on the results we obtained to answer \RQ{2}, to eliminate smell types that are unlikely to be related with bug-proneness, we created several sets of smell types by excluding those with a lower relative risk and retain only those with a higher relative risk with different boundaries.
Finally, we compare the performance of the technique by specifying these five settings.
\begin{itemize}
  \item \CP{$s_1$}{all smell types}~(16): all types of smells,
  \item \CP{$s_2$}{rare selected smell types}~(14): types of smells whose relative risk is greater than 1,
  \item \CP{$s_3$}{medium rare selected smell types}~(12): types of smells whose risk is greater than that of all types of smells~(5.735\%),
  \item \CP{$s_4$}{medium selected smell types}~(10): types of smells whose relative risk is greater than that of all types of smells~(2.785),
  \item \CP{$s_5$}{well selected smell types}~(5): top five types of smells regarding their relative risk; their relative risk is greater than 4.
\end{itemize}
The concrete types are specified at the left of Table~\ref{t:smell_ratio}.

To compare all configurations, as discussed earlier, we applied our technique to all 150~(= 3 granularity levels $\times$ 10 aggregators $\times$ 5 type selectors) configurations and calculated the accuracy.

In addition to the five selectors, we also prepared special selectors that use only one smell type to investigate the performance of each smell type.
When selecting these selectors together with other perspectives,
  1) the smell granularity is automatically assigned to a specific value according to the type,
  2) nested aggregators~($a_7$--$a_{10}$) are unnecessary because the first aggregation step results in only one value instance, and these aggregators produce the same result as $a_2$ or $a_4$, and
  3) For class-level smell types, severity-based aggregators~($a_1$, $a_2$, $a_5$, and $a_6$) and count-based ones~($a_3$ and $a_4$) will respectively produce the same result because only one smell instance is assumed to be detected, and only the aggregators of $a_2$ and $a_3$ are enough to be considered as their representatives.
Therefore, we prepared $1 \times 2 \times 7\text{~(class-level)} + 1 \times 6 \times 9\text{~(method-level)} = 68$ configurations.
We compare the best configuration of each smell type to determine which smell types contributed to the performance improvements.

\subsubsection{Results}

The accuracy for the top 20 configurations~(\rnum{1} to \rnum{20}) is listed in Table~\ref{t:conf-perf}, sorted in descending order of MAP.
Note that two special configurations were additionally included in the table.
\rnum{74} is the configuration equivalent to that used in our previous study.
\rnum{0} is a pseudo-ideal configuration that allows a different configuration selection for each system as though we could know the best configuration for each system.
The annotated numbers with parentheses highlights represent the rank of each metric value for the top three configurations.
In addition, values at the column ``\# systems'' indicate the numbers of systems that the smell-aware approach outperformed the baseline bug localization technique, i.e., the cases where $\alpha > 0$ was used.
As we can see, different configurations produce different numbers of systems to succeed, mainly depending on the smell types to be used.

Considering the case of Top 1, Top 5, MRR, and MAP, the best configuration that yields the best performance was the configuration \rnum{1} \Config{\CP{$g_3$}{both class and method levels}, \CP{$a_3$}{existence of smells}, \CP{$s_5$}{well selected smell types}}.
On the other hand, when considering Top 10, the combination \rnum{2} \Config{\CP{$g_3$}{both class and method levels}, \CP{$a_2$}{maximum severity}, \CP{$s_3$}{medium rare selected smell types}} performed the best.
In terms of overall performance, \rnum{1}, \rnum{2}, and \rnum{3} performed well.
Based on these observations, providing the technique with an appropriate configuration enables it to significantly outperform the technique developed in our previous study.

\begin{figure}[tb]\centering
	\includegraphics[width=0.7\linewidth]{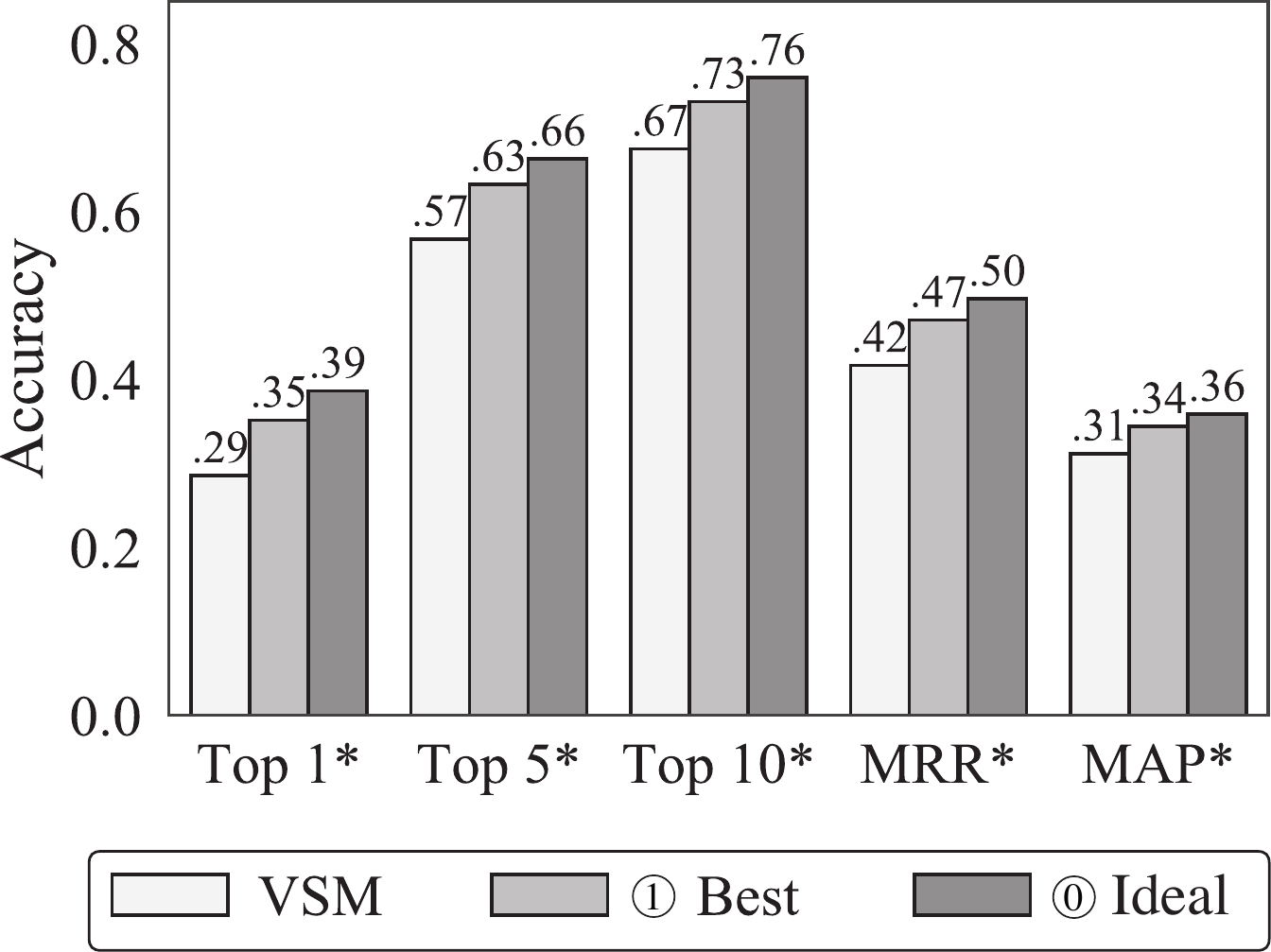}
	\caption{Accuracy comparison of smell-aware bug localization with the best configuration and \BLT{VSM}.}\label{f:best-vsm}
\end{figure}

\begin{figure}[tb]\centering
	\newcommand{\figheight}{6cm}
	\begin{subfigure}{\linewidth}\centering
		\includegraphics[scale=0.4]{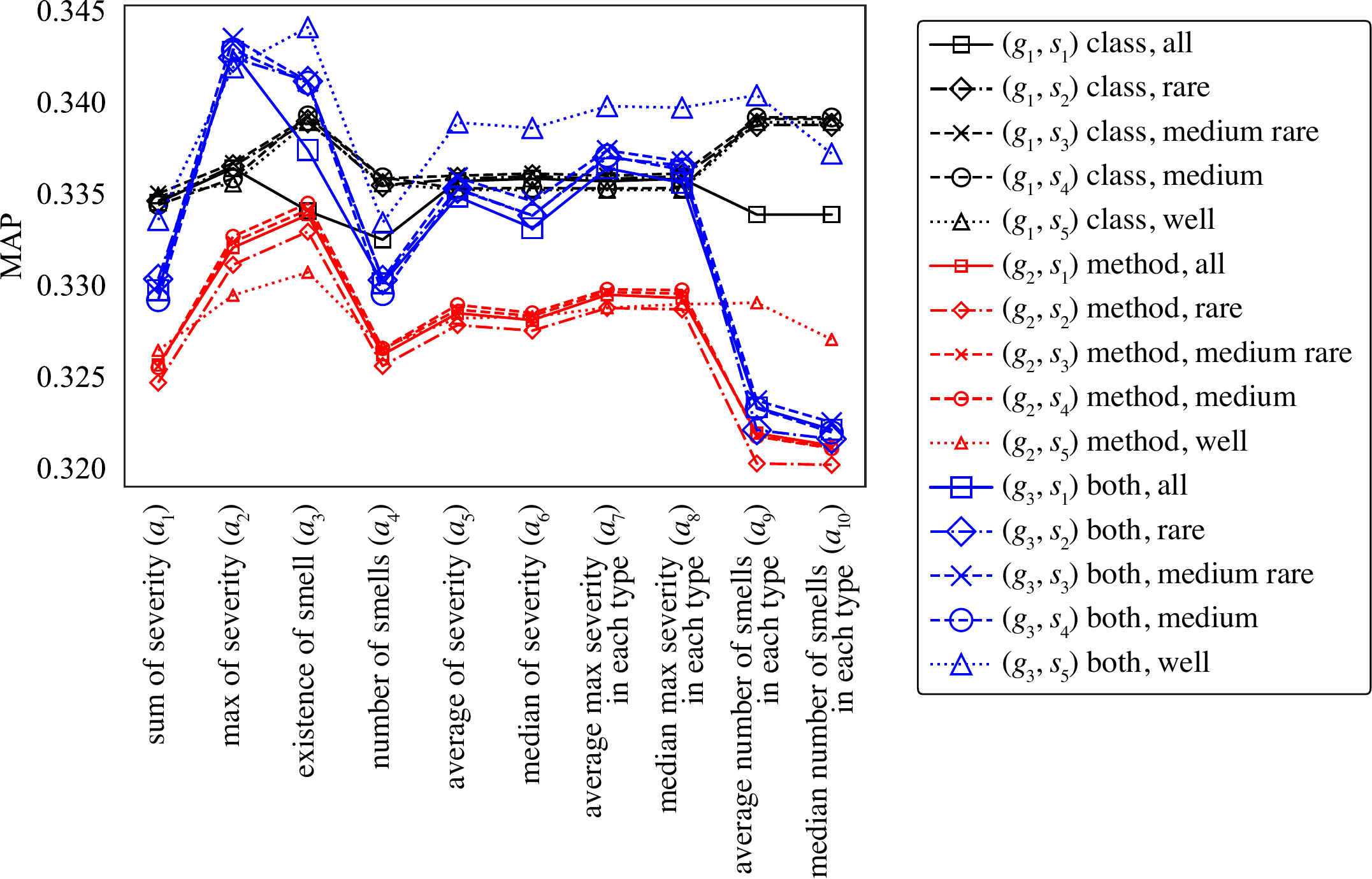}
		\caption{Smell granularity and selector.}\label{f:perspective-gs}
	\end{subfigure} \\ \vspace{1em}
	\begin{subfigure}{\linewidth}\centering
		\includegraphics[scale=0.4]{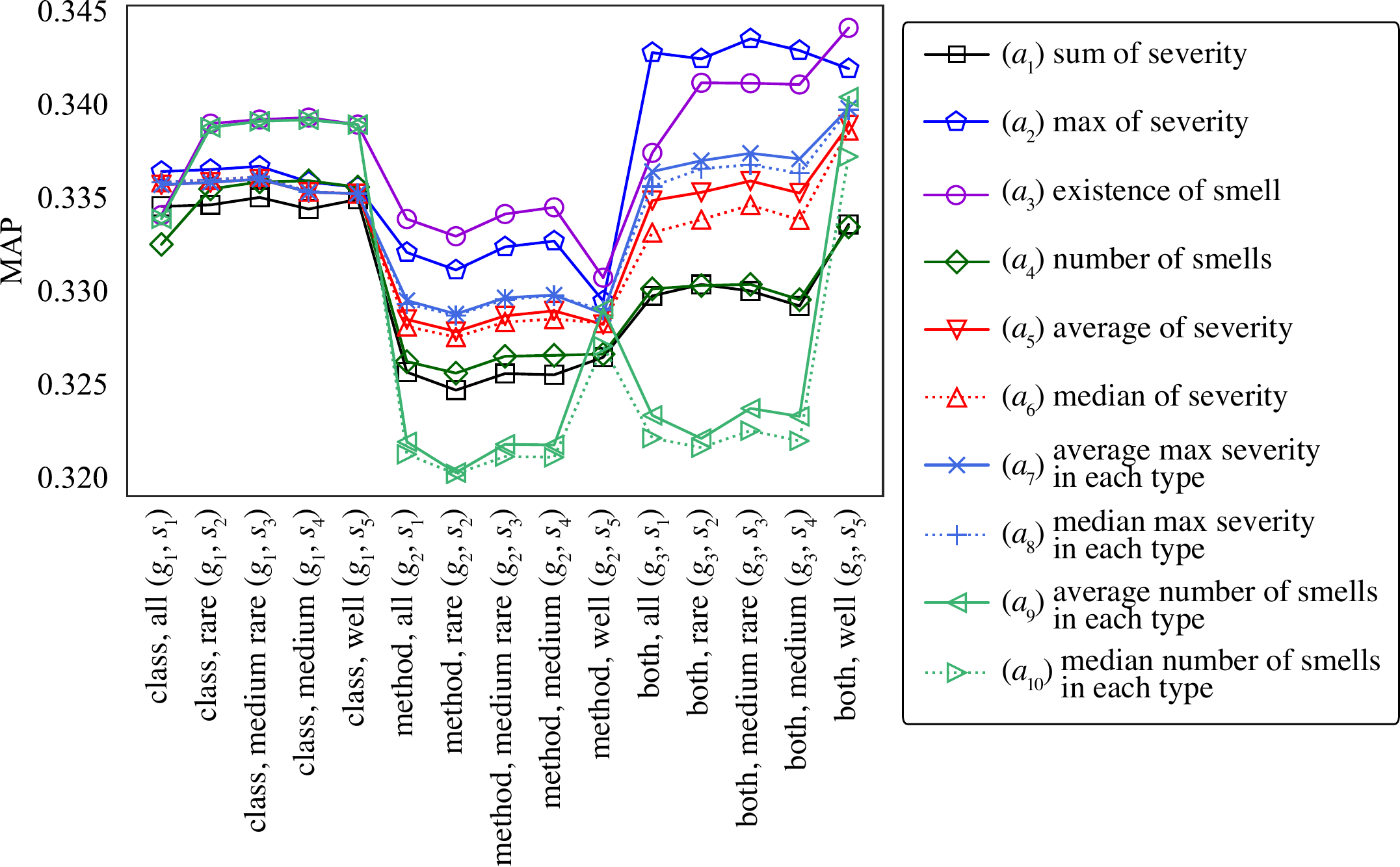}
		\caption{Smell aggregator.}\label{f:perspective-a}
	\end{subfigure}
	\caption{MAP performance of configurations of specific perspectives across the other perspectives.}\label{f:perspective}
\end{figure}

Furthermore, to address \RQ{1}, we used the configuration \rnum{74} to conduct the experiment and found the difference in terms of MAP to be statistically insignificant, as shown in Fig.~\ref{f:orig-vsm}.
Nevertheless, when we reran the experiment with the configuration \rnum{1}, we not only observed statistically significant results in all metrics~($p<0.01$; Cliff's delta: 0.066, 0.065, 0.056, 0.066, and 0.054, all \emph{negligible}), but also an increase in the improvement of all metrics, as shown in Fig.~\ref{f:best-vsm}.
For example, Top 10 increased by 0.056~(from 0.675 to 0.731), which means that the total number of bug reports with gold modules in their top 10-ranked items increased by 391~(from 4,676 to 5,067).
This figure also presents the results obtained using the ideal configuration \rnum{0}, which show additional improvements compared to \rnum{1}.
This result indicates that there remains a scope for improvement when using smell-aware bug localization if we know the best smell configuration per system.

\textbf{Analysis on each configuration parameter.}
We compared the MAP scores according to each configuration parameter to determine the contribution of each parameter to the performance.
The following discussion is based on comparisons of the MAP results by fixing all perspectives other than the one to be discussed.
Figure~\ref{f:perspective} shows the difference in the MAP performance of specific perspectives over the other perspectives.
Different combinations of smell granularity~($g$) and smell selector~($s$) are compared in Fig.~\ref{f:perspective-gs}, whereas the smell aggregators are compared in Fig.~\ref{f:perspective-a}.

The three different colors in Fig.~\ref{f:perspective-gs} specify the smell granularity that was used.
The results in the figure show that the configurations using \CP{$g_1$}{class level} and \CP{$g_3$}{both class and method levels} had higher scores depending on the used aggregators, and \CP{$g_2$}{method level} produced worse results in general.
In particular, the use of both level smells produced more accurate results when using it with \CP{$a_2$}{maximum severity} or \CP{$a_3$}{existence of smells}.
This result suggests that adding method-level smells in addition to class-level smells may be effective in extending the range of smells to be used, but considering all of them may also have a negative effect.
For instance, \CP{$a_1$}{sum of severity} and \CP{$a_4$}{number of smells} add up all smells as equivalent, even if class-level smells exist.
This kind of smell usage was not effective because method-level smells may decrease the importance of class-level smells.
In contrast, \CP{$a_2$}{maximum severity} and \CP{$a_3$}{existence of smells} are considered to be effective because they can consider method-level smells when class-level smells do not exist or when the method-level smells have higher severity than the class-level smells.

In the smell aggregators shown in Fig.~\ref{f:perspective-a}, the configurations using \CP{$a_2$}{maximum severity} and \CP{$a_3$}{existence of smells} produced higher values.
Other types of aggregators produced worse results in general, except for the average or median of the number of smells in each type~($a_9$ and $a_{10}$) with a certain level of selection in class-level smells~($g_1$, $s_2$--$s_5$).
In addition, we found that configurations using median~($a_6$, $a_8$, and $a_{10}$) produced very similar results to those using average~($a_5$, $a_7$, and $a_9$).
These results suggest that indicators such as \CP{$a_2$}{maximum severity} and \CP{$a_3$}{existence of smells} should be used.

For the smell selectors, as shown in Fig.~\ref{f:perspective-a}, the use of selection~($s_2$--$s_5$) tended to produce more accurate results than the configurations using \CP{$s_1$}{all smell types}.
In particular, when using \CP{$g_3$}{both class and method levels}, the use of \CP{$a_2$}{maximum severity} was the best choice at a certain level of selection, whereas the use of \CP{$a_3$}{existence of smells} was more effective if it was used together with \CP{$s_5$}{well selected smell types}.
This result suggests the effectiveness of smell selection based on the likelihood of containing bugs, as indicated in Table~\ref{t:smell_ratio} in general.
In addition, the use of the severity degree tends to be more effective if a broader range of smell types are used.

\begin{table}[tb]\centering
  \caption{Accuracy of each configuration when only one smell type is used}\label{t:one-type}
  {\tabcolsep=3.5pt\scriptsize\begin{tabular}{llrrrrr} \toprule
          $s$: Smell type & $a$: Aggregator  & Top 1  & Top 5  & Top 10 & MRR    & MAP \\ \midrule
                \GodClass & \CP{$a_3$}{existence} & 0.3310 & 0.6148 & 0.7162 & 0.4556 & 0.3349 \\
           \BlobOperation & \CP{$a_3$}{existence} & 0.3311 & 0.6045 & 0.7091 & 0.4499 & 0.3293 \\
               \BlobClass & \CP{$a_2$}{severity}  & 0.3073 & 0.5914 & 0.6951 & 0.4343 & 0.3218 \\
             \FeatureEnvy & \CP{$a_3$}{existence} & 0.3086 & 0.5902 & 0.6940 & 0.4328 & 0.3208 \\
              \DataClumps & \CP{$a_3$}{existence} & 0.3052 & 0.5924 & 0.6963 & 0.4332 & 0.3204 \\
     \InternalDuplication & \CP{$a_3$}{existence} & 0.2985 & 0.5823 & 0.6884 & 0.4262 & 0.3179 \\
       \IntensiveCoupling & \CP{$a_3$}{existence} & 0.2966 & 0.5820 & 0.6873 & 0.4253 & 0.3163 \\
      \SchizophrenicClass & \CP{$a_3$}{existence} & 0.2955 & 0.5747 & 0.6811 & 0.4230 & 0.3148 \\
     \ExternalDuplication & \CP{$a_3$}{existence} & 0.2919 & 0.5741 & 0.6814 & 0.4212 & 0.3144 \\
           \MessageChains & \CP{$a_3$}{existence} & 0.2919 & 0.5754 & 0.6829 & 0.4217 & 0.3140 \\
    \RefusedParentBequest & \CP{$a_3$}{existence} & 0.2927 & 0.5726 & 0.6813 & 0.4215 & 0.3139 \\
          \ShotgunSurgery & \CP{$a_3$}{existence} & 0.2899 & 0.5726 & 0.6796 & 0.4203 & 0.3138 \\
      \SiblingDuplication & \CP{$a_3$}{existence} & 0.2916 & 0.5719 & 0.6791 & 0.4205 & 0.3134 \\
               \DataClass & \CP{$a_3$}{existence} & 0.2929 & 0.5728 & 0.6801 & 0.4210 & 0.3131 \\
        \TraditionBreaker & \CP{$a_3$}{existence} & 0.2880 & 0.5699 & 0.6768 & 0.4188 & 0.3124 \\
      \DistortedHierarchy & \CP{$a_3$}{existence} & 0.2855 & 0.5672 & 0.6747 & 0.4169 & 0.3117 \\\bottomrule
  \end{tabular}}
\end{table}

\textbf{Analysis of individual smell types.}
The best configurations when their selectors use only one smell type are presented in Table~\ref{t:one-type}.
Each row in this table indicates the performance of the best configuration when a specific smell type is used as its selector.
Rows are ordered by their MAP score.
Although several smell types, such as \GodClass, \BlobOperation, or \BlobClass, outperformed other smell types, no one outperformed the best configurations in Table~\ref{t:conf-perf}.
This result shows that an awareness of multiple smell types improves the performance to a greater extent than only one specific smell type.

\begin{figure}[tb]\centering
	\includegraphics[width=\linewidth]{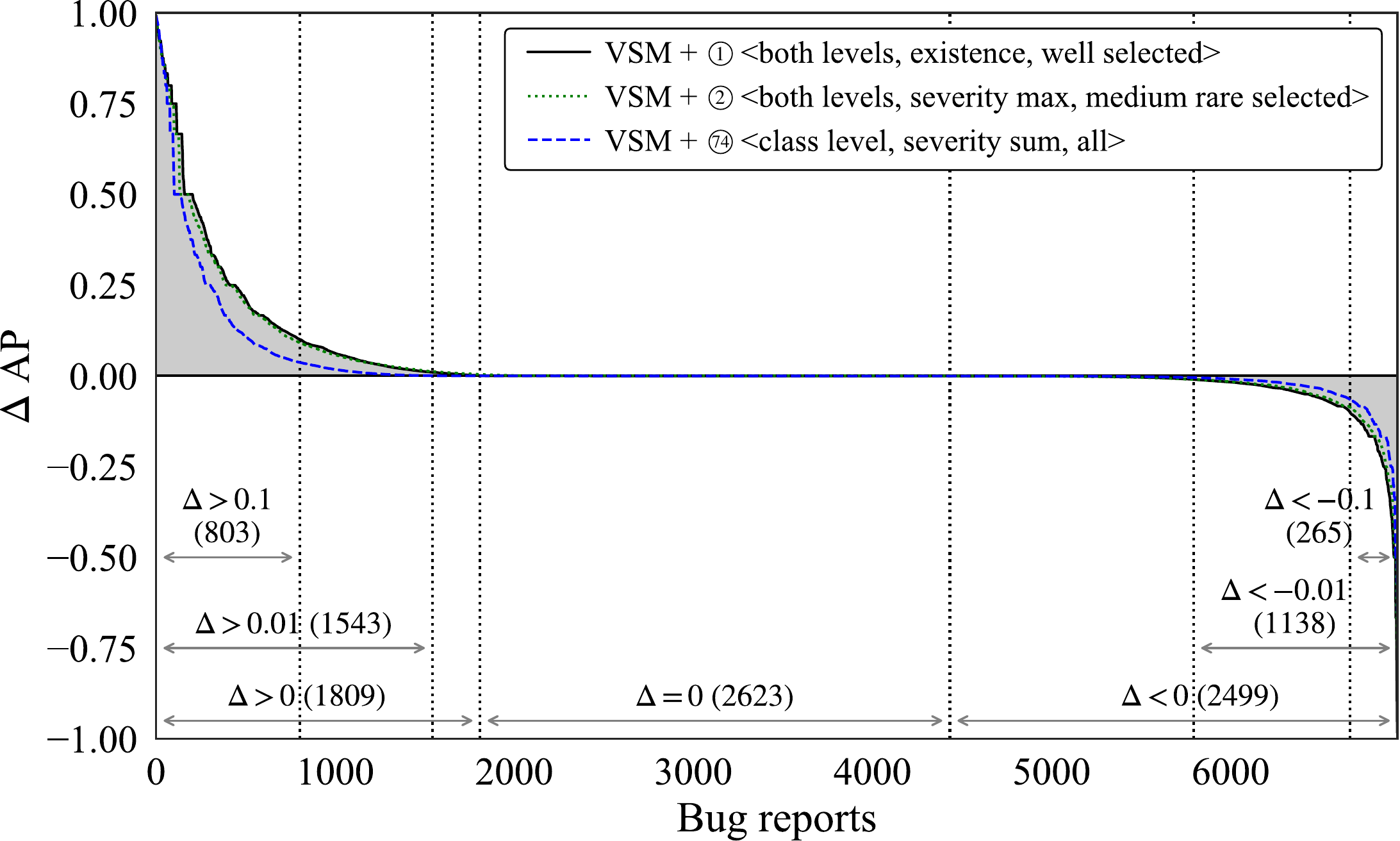}
	\caption{Distribution of AP improvements.}\label{f:ap-dist}
\end{figure}

\begin{figure*}[tb]\centering
	\newcommand{\figheight}{2.6cm}
  {\scriptsize\tabcolsep=0.1em\begin{tabular}{ccccc}
    \includegraphics[height=\figheight]{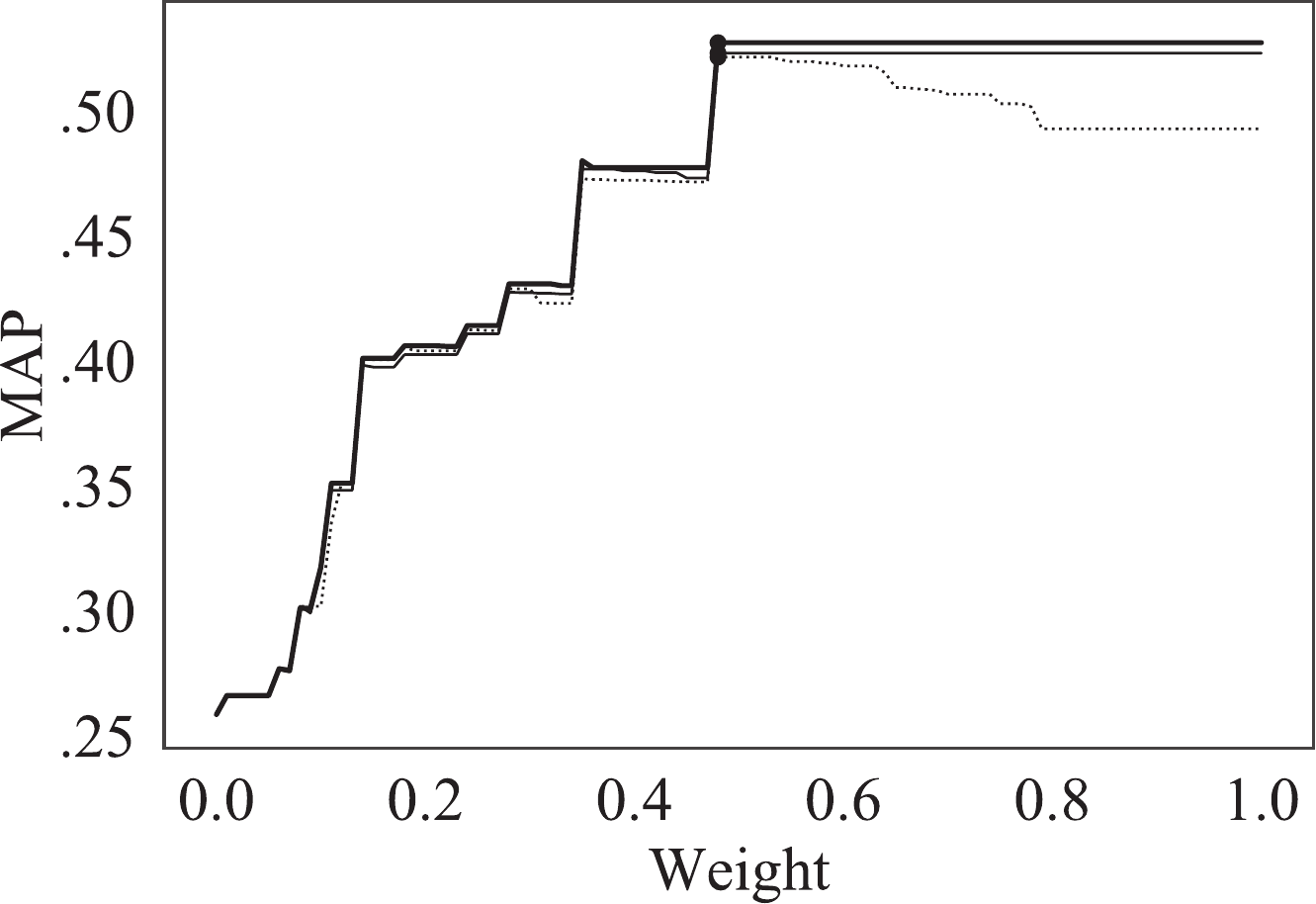} &
    \includegraphics[height=\figheight]{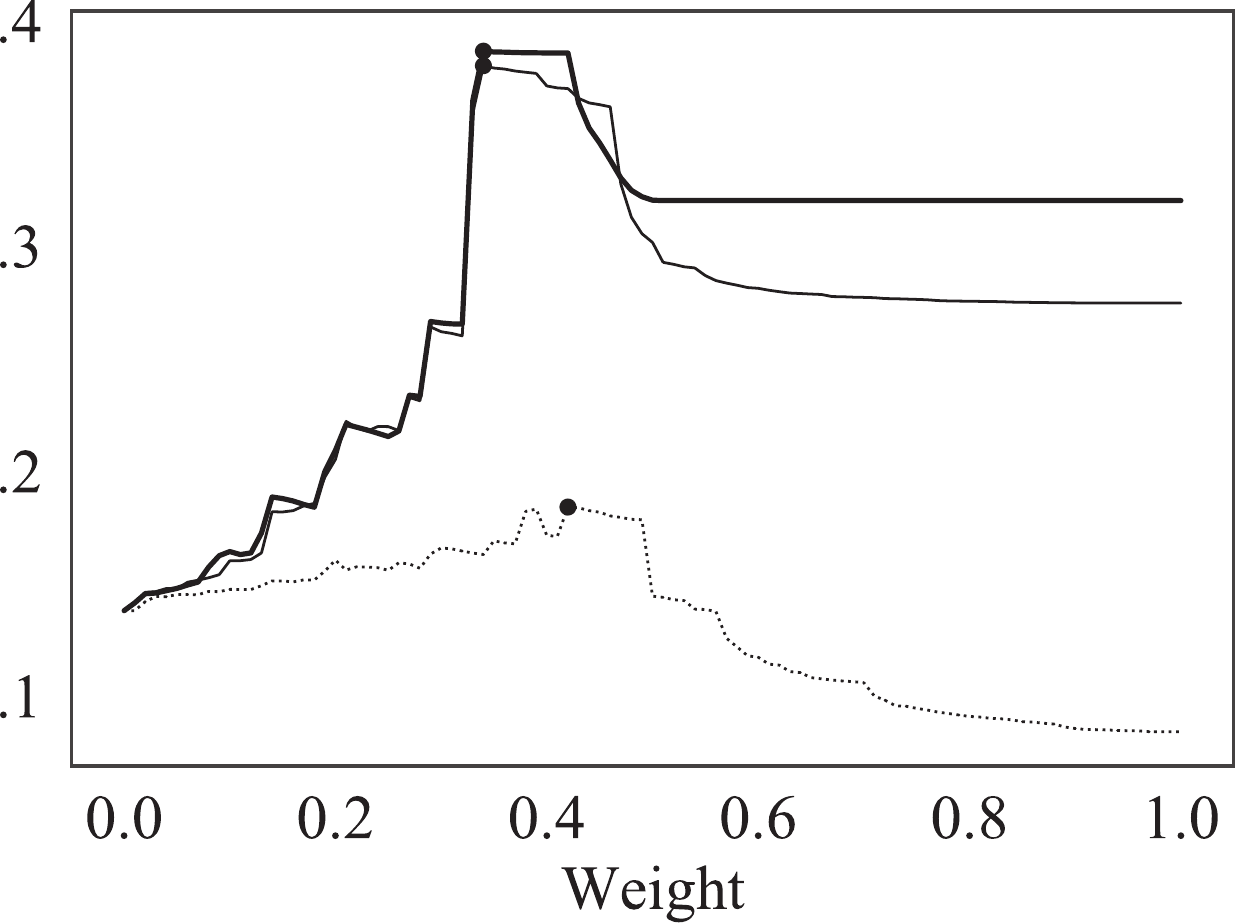} &
    \includegraphics[height=\figheight]{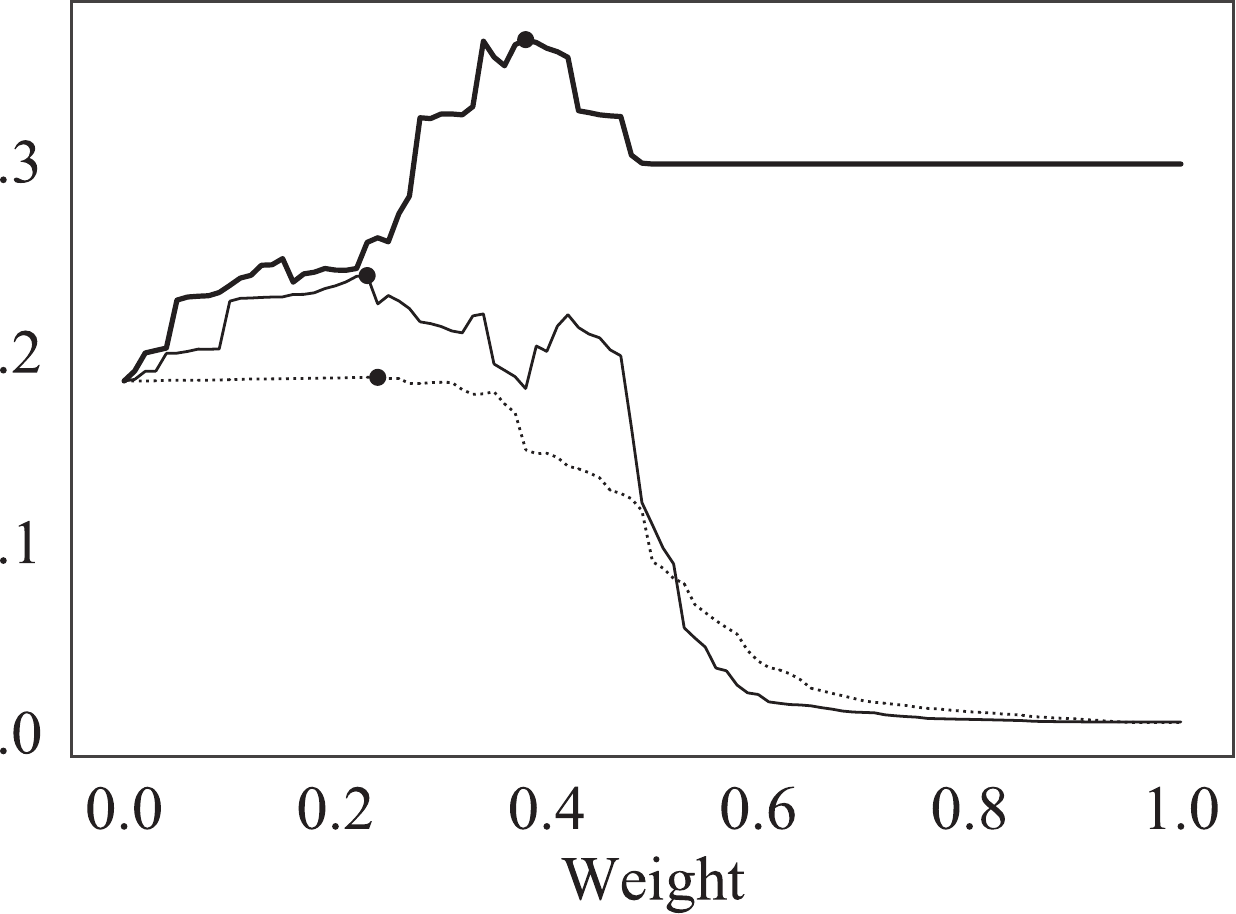} &
    \includegraphics[height=\figheight]{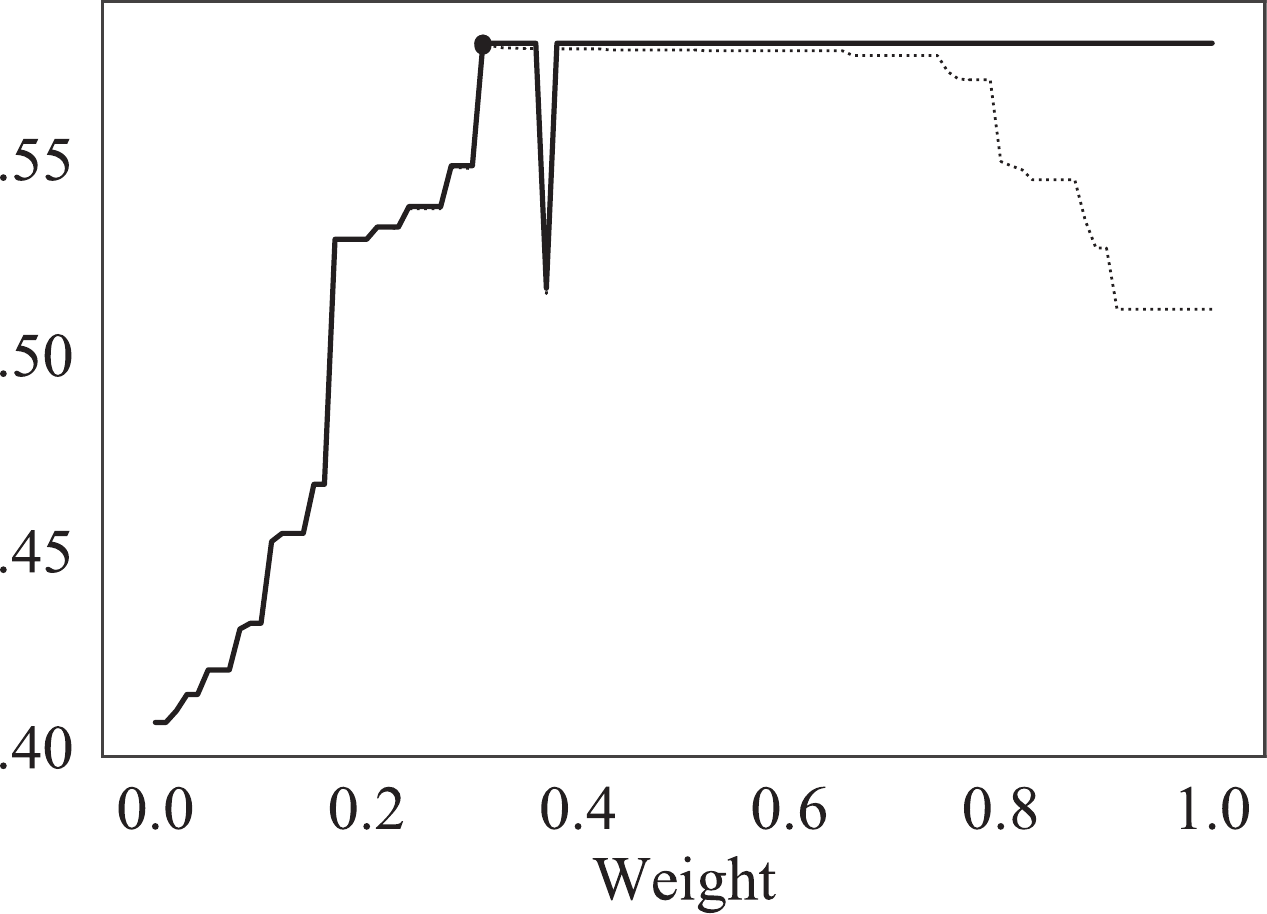} &
    \includegraphics[height=\figheight]{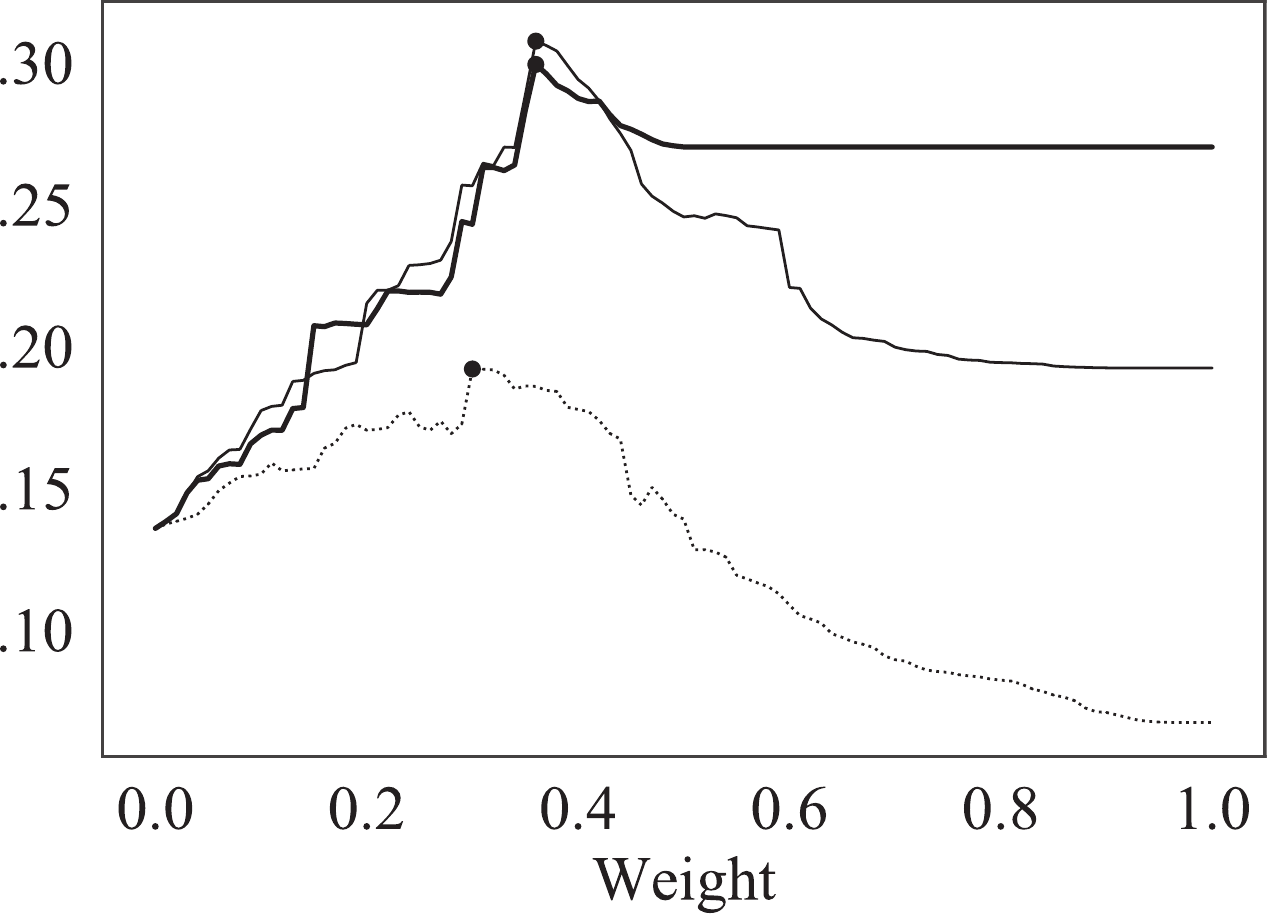} \\
    (a) \textsf{DATAMONGO 1.1.2}~(1st) &
    (b) \textsf{HBASE 0.98.19}~(2nd) &
    (c) \textsf{CAMEL 2.14.2}~(5th) &
    (d) \textsf{DATAREST 1.0.0}~(6th) &
    (e) \textsf{HIVE 0.11.0}~(7th) \\ ~ \\
    \multicolumn{5}{c}{\includegraphics[width=15cm]{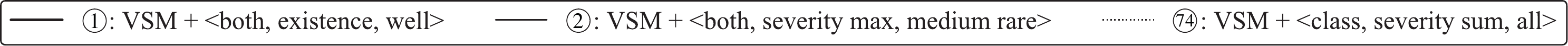}}
  \end{tabular}}
  \caption{Distribution of MAP values according to $\alpha$ parameter.}\label{f:system-plots}
\end{figure*}

\textbf{Distribution of AP improvements for each bug report.}
To improve our understanding of the performance improvement, we analyzed the distribution of AP improvements of 6,931 bug reports when applying smell-aware bug localization to the \BLT{VSM} results.
The distribution is visualized in Fig.~\ref{f:ap-dist}.
In this figure, the black line is plotted to express the AP deltas~($\Delta \mathit{AP} = \mathit{AP}_\text{smell-aware} - \mathit{AP}_\mathsf{VSM}$) in descending order, obtained using the best configuration \rnum{1}.
As is shown, the smell-aware bug localization improved the overall accuracy because the total improvements~(top left) exceeded the total decreases~(bottom right).
However, the use of the smell-aware bug localization did not improve the values of all bug reports; instead, certain values were less accurate.
In this result, out of 6,931 bug reports, 1,809 of AP increased, whereas 2,499 decreased.
However, in most of them, the value of delta was small, and for those with an absolute delta greater than 0.01, 1,543 increased, whereas 1,138 decreased.
In particular, 803 reports improved and 265 did not improve when the absolute delta was greater than 0.1.
We consider that the accuracy to have improved because the number of bug reports with large improvements was relatively more extensive than that with large decreases.
The figure also includes plots of the original configuration \rnum{74} and the second-best configuration \rnum{2}.
Clearly, the degree of improvement by the best configuration is much higher than that of the original configuration.
Moreover, the difference with the second-best configuration is small.

\textbf{Benefited and non-benefited systems.}
Figure~\ref{f:system-plots} shows the results of the top systems with the biggest improvement in MAP when using the best configuration \rnum{1}.
We picked up first, second, fifth, sixth, and seventh-top systems because they are the top systems when picking up only one version from the same project.
In the figure, each graph shows the MAP value at different $\alpha$ for each system.
For comparison, the results of the two best configurations as well as the original configuration, i.e., \rnum{1}, \rnum{2}, and \rnum{74}, are plotted.
The point for $\alpha = 0$ refers to the accuracy when using only the baseline IR-based bug localization technique, i.e., \BLT{VSM}, whereas $\alpha = 1$ refers to the accuracy when using only the code smell property to localize bugs.
We can see two typical shapes in these plots:
\begin{itemize}
  \item \textit{Mountain-shaped}:
  A certain level of blending of IR-based bug localization and the smell information gave the best result, i.e., $0 \not\in A \wedge 1 \not\in A$, where $A$ is the set of best $\alpha$ values for the setting.
  Including \textsf{HBASE 0.98.19}, \textsf{CAMEL 2.14.2}, and \textsf{HIVE 0.11.0},
  the plots of 196 and 225 out of 309 systems in \rnum{1} and \rnum{2} follow this style, respectively.
  \item \textit{Plateau-shaped}:
  When $\alpha$ exceeds a certain value, MAP stayed at the highest value, i.e., $0 \not\in A \wedge 1 \in A$.
  This case shows that smells caught the buggy location without the help of the textual similarity of the IR-based bug localization.
  Including \textsf{DATAMONGO 1.1.2} and \textsf{DATAREST 1.0.0}, the plots of 5 and 4 out of 309 systems in \rnum{1} and \rnum{2} follow this style, respectively.
\end{itemize}
Also, we can see that the best configuration did not always lead to the best result; for example, in Fig.~\ref{f:system-plots}e, the second-best configuration led to the best result.
Note that 101 of 309 systems used the parameter of $\alpha = 0$ when using the best configuration \rnum{1}.
The most typical case for this situation is simply that the buggy portions to be fixed were not smelly.

\textbf{In conclusion, the configurations using \textit{both class and method levels} for the granularity yielded the best result.
In terms of the combinations of the aggregator and the selector, the \textit{existence} of \textit{well selected types of smells} or \textit{maximum severity} of \textit{medium rare selected types of smells} yielded the best results.}

\subsection{\RQ{4}: Is the performance of smell-aware bug localization superior to that of state-of-the-art bug localization techniques?}\label{ss:rq4}

\subsubsection{Motivation}

When addressing \RQ{1} and \RQ{3}, we found that the smell-aware bug localization technique can improve the performance when combined with the \BLT{VSM} technique.
However, many bug localization techniques have been proposed to improve the \BLT{VSM} technique such as \BLT{rVSM}, \BLT{BugLocator}, \BLT{BRTracer}, \BLT{BLUiR}, and \BLT{AmaLgam}.
The goal of this study is to verify whether the smell-aware bug localization technique can also be used to improve bug localization techniques other than \BLT{VSM}.

\subsubsection{Study Design}

We implemented the smell-aware bug localization technique using six existing bug localization techniques as baselines.
Specifically, we used the output score of each technique as $\nScore$ in Section~\ref{s:technique}.
For $\nSmell$, we apply the best configurations discussed in \RQ{3}, that is, \rnum{1} and \rnum{2} in Table~\ref{t:conf-perf}.

Finally, we compare the accuracy of the ranking produced using gBLI and each baseline technique.

\subsubsection{Results}

\begin{figure*}[tb]\centering
	\newcommand{\figheight}{3.8cm}
	\begin{subfigure}{0.4\linewidth}\centering
		\includegraphics[height=\figheight]{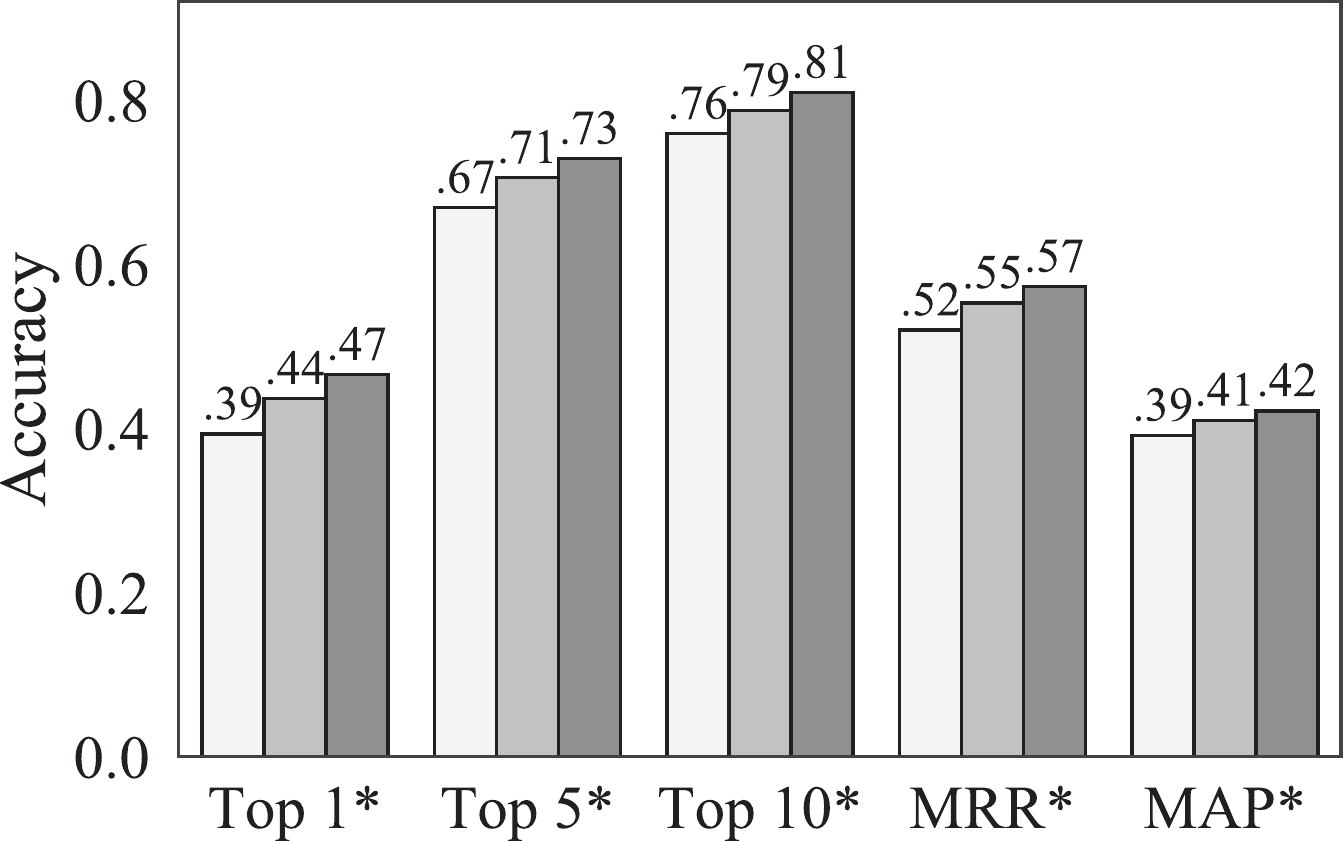}
		\caption{\BLT{rVSM}.}\label{f:best-rvsm}
	\end{subfigure}
	\begin{subfigure}{0.4\linewidth}\centering
		\includegraphics[height=\figheight]{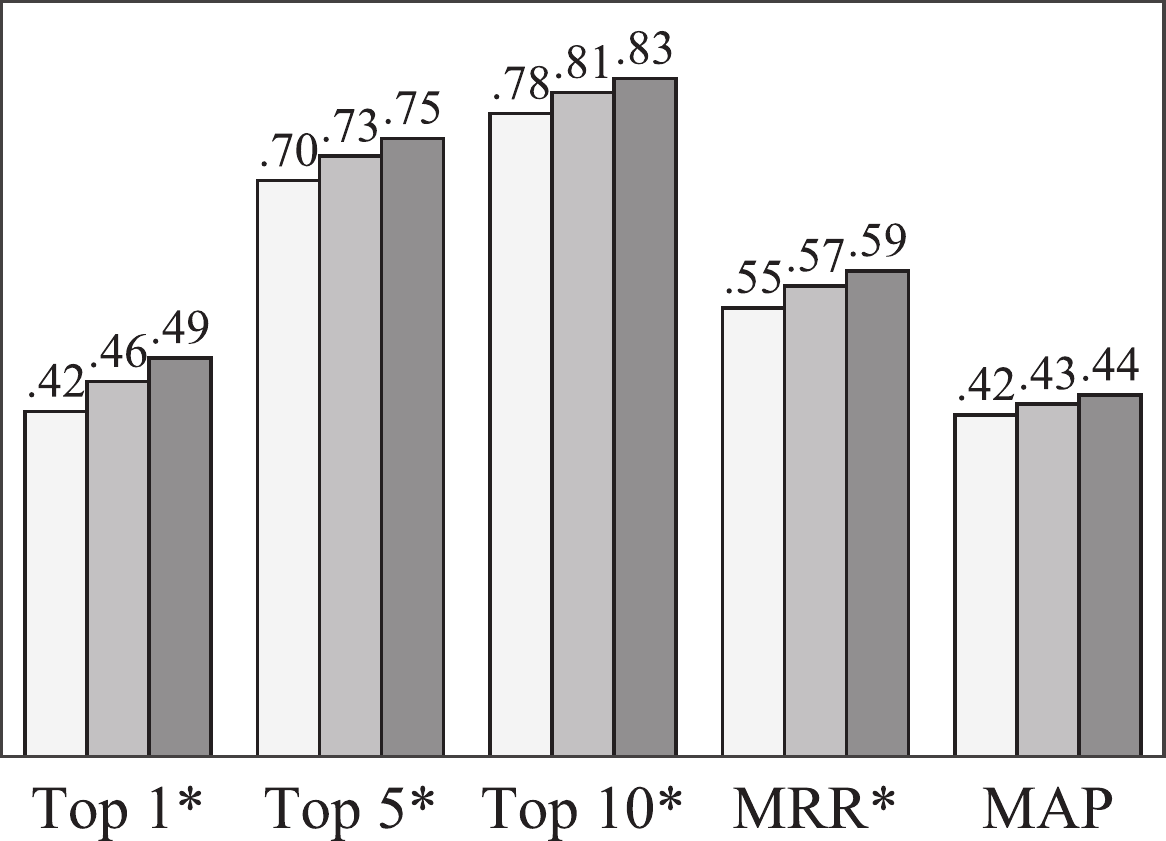}
		\caption{\BLT{BugLocator}.}\label{f:best-buglocator}
	\end{subfigure}
	\\\vspace{3mm}
	\begin{subfigure}{0.355\linewidth}\centering
		\includegraphics[height=\figheight]{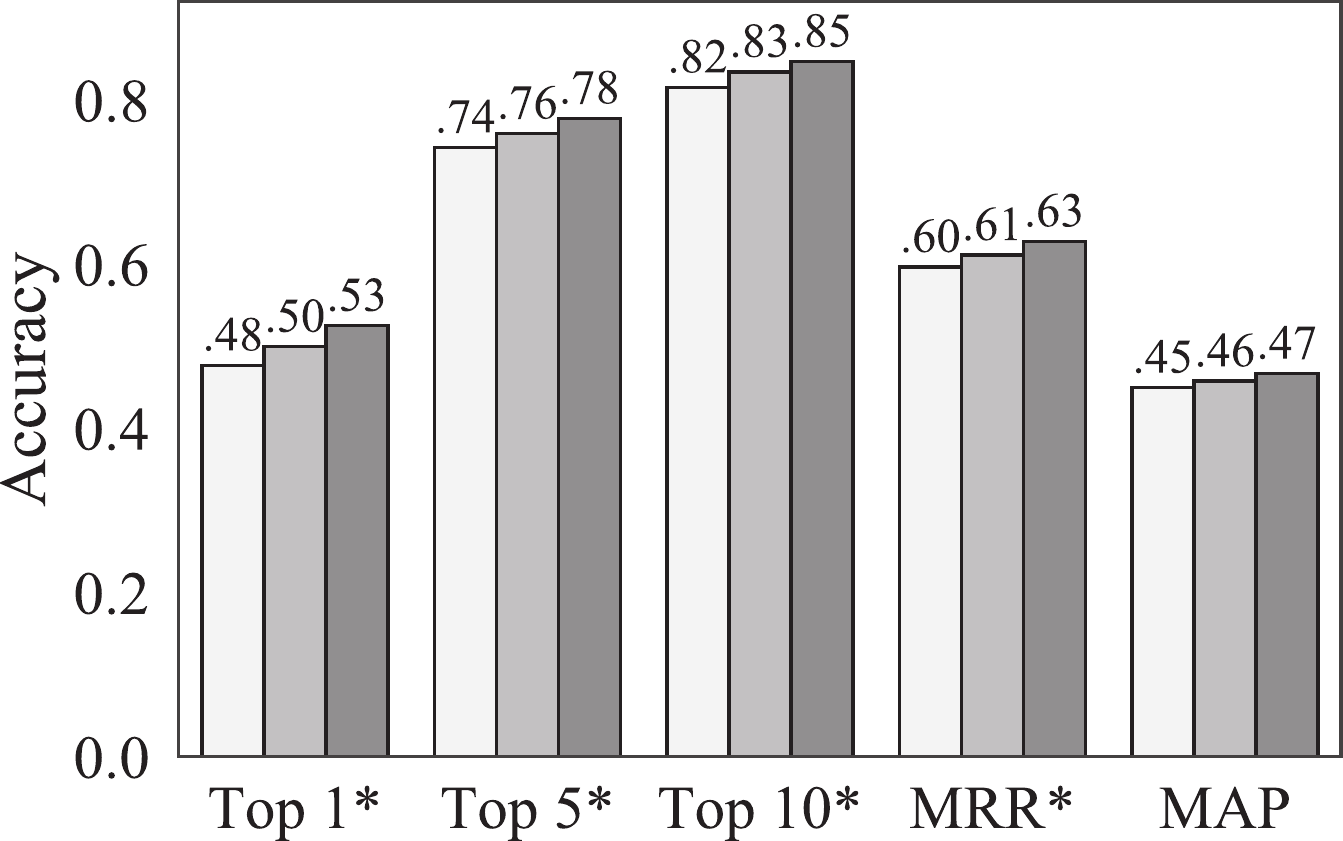}
		\caption{\BLT{BRTracer}.}\label{f:best-brtracer}
	\end{subfigure}
	\begin{subfigure}{0.31\linewidth}\centering
		\includegraphics[height=\figheight]{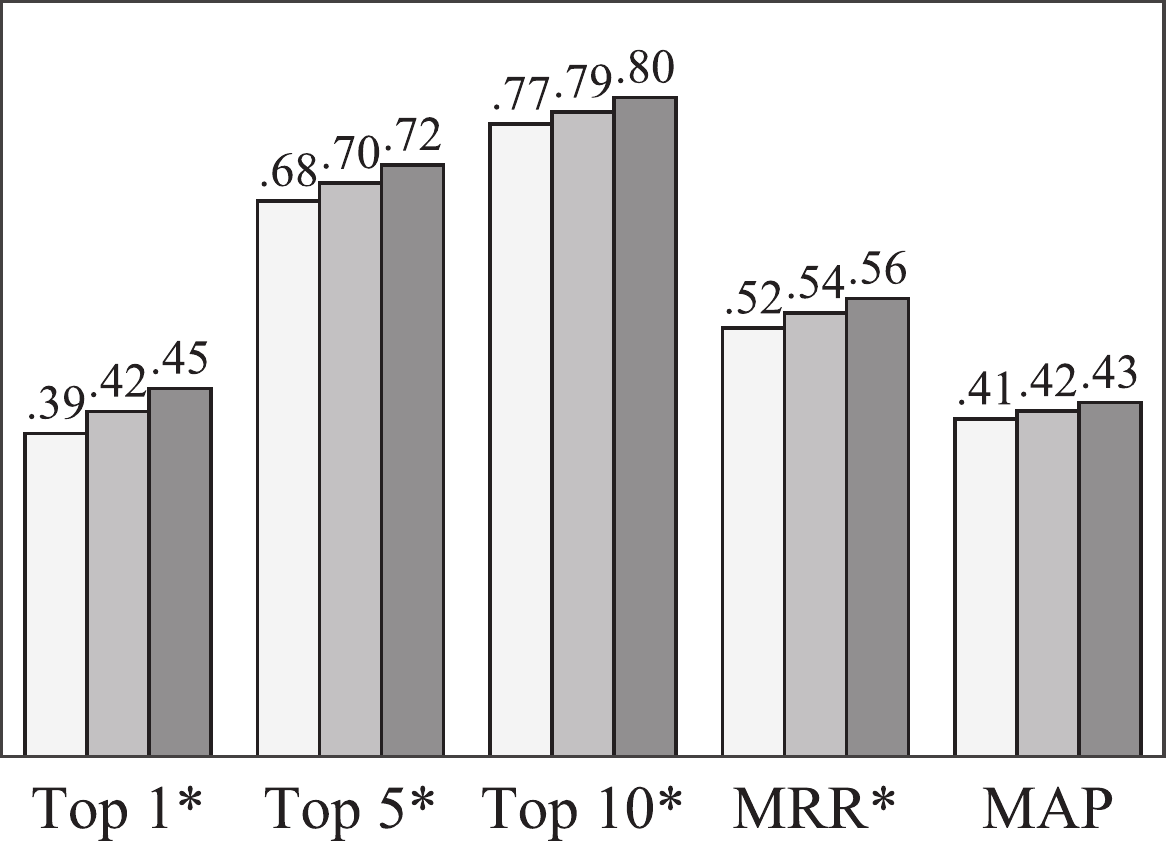}
		\caption{\BLT{BLUiR}.}\label{f:best-bluir}
	\end{subfigure}
	\begin{subfigure}{0.31\linewidth}\centering
		\includegraphics[height=\figheight]{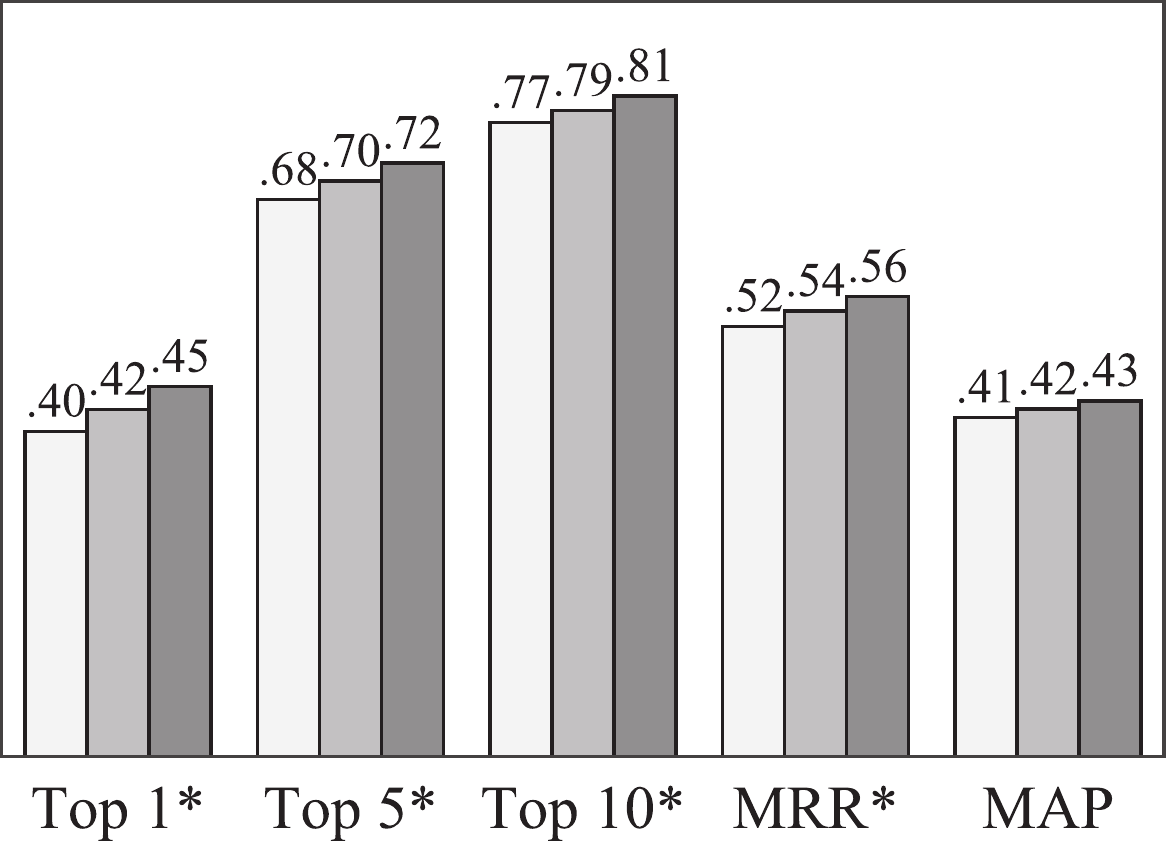}
		\caption{\BLT{AmaLgam}.}\label{f:best-amalgam}
	\end{subfigure}
	\\\vspace{1mm}
	\begin{subfigure}{\linewidth}\centering
		\includegraphics[height=0.45cm]{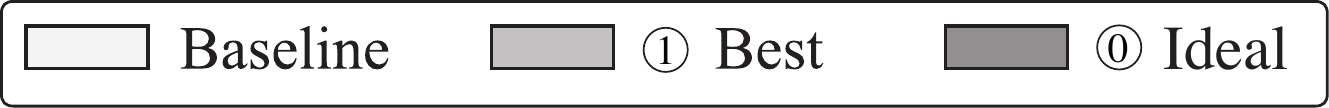}
	\end{subfigure}
	\caption{Accuracy of the smell-aware bug localization technique relative to that of the baseline.}\label{f:best-results}
\end{figure*}

The results of each technique using with \rnum{1}~\Config{\CP{$g_3$}{both class and method levels}, \CP{$a_3$}{existence of smells}, \CP{$s_5$}{well selected smell types}} as the smell configuration are shown in Fig.~\ref{f:best-results}.
Although we applied two configurations~(\rnum{1} and \rnum{2}), we reached almost the same conclusion for each configuration.
To save space, we mainly selected \rnum{1} to explain the details and clarified major differences if exist.

Figure~\ref{f:best-rvsm} shows the result of the \BLT{rVSM} technique.
The results obtained with the smell-aware bug localization technique are an improvement relative to the baseline by approximately 11.0\%, 5.4\%, 3.7\%, 6.4\%, and 4.7\% in relative comparison~(0.043, 0.036, 0.028, 0.033, and 0.018 in absolute comparison), for Top 1, Top 5, Top 10, MRR, and MAP, respectively.
All of these improvements are statistically significant~(Cliff's delta: 0.043, 0.036, 0.028, 0.041, and 0.030, all \emph{negligible}).
This indicates that smell information is useful even with a bug localization technique that uses information about the size of the source code.
This result suggests that the smell-aware bug localization technique is effective not only because of the size of the source code but also because of other factors.

In the case of \BLT{BugLocator}, as shown in Fig.~\ref{f:best-buglocator}, a similar result was observed.
All of the improvements except for MAP are statistically significant: 8.6\%, 4.3\%, 3.2\%, 4.8\%, and 3.1\% in relative comparison~(0.036, 0.030, 0.025, 0.026, and 0.013 in absolute comparison) for Top 1, Top 5, Top 10, MRR, and MAP~(Cliff's delta: 0.036, 0.030, 0.025, 0.033, and 0.022, all \emph{negligible}), respectively.
Note that the MAP difference was statistically significant when using the configuration \rnum{2}.
This result suggests that our technique even improves the technique that used information about bug reports in the past.

However, in the cases of \BLT{BRTracer}, \BLT{BLUiR}, and \BLT{AmaLgam} in Figs.~\ref{f:best-brtracer}, \ref{f:best-bluir}, and \ref{f:best-amalgam}, statistically significant improvements in all the metrics except MAP are obvious.
Specifically, the improvement in MAP in \BLT{BRTracer} is only 1.7\%, which is the lowest improvement among all the projects.

\textbf{In conclusion, optimization of the configuration of the smell-aware bug localization technique can improve state-of-the-art bug localization techniques.}


\section{Threats to Validity}\label{s:threats_to_validity}

\subsection{Internal Validity}\label{ss:internal-validity}

In this study, the weight $\alpha$ was assigned the optimal value to maximize each accuracy metric of each version.
This is intended to avoid the possibility of not being able to observe the effect of smells because of the effect on the accuracy as a result of the choice of $\alpha$.
However, in \RQ{1} and \RQ{4}, we also discussed the extent to which the smell-aware bug localization technique is superior to the baseline.
Therefore, the threat that remains is that the used optimal $\alpha$ values differ from the practical values of $\alpha$.
In particular, the possibility of such cases occurring in versions with a small number of bug reports is high because the values of $\alpha$ might be biased toward those bug reports. 
Therefore, we conducted our experiments by excluding versions with fewer than five bug reports.
This exclusion prevented the value of $\alpha$ from being over-optimized in a version with fewer bug reports.

Another threat is the accuracy of the output of Bench4BL, which we used in our study.
As mentioned in Section~\ref{sss:selection}, we mitigated this threat by excluding inconsistent results from the output of Bench4BL.
However, the validity of the output depends on the quality of Bench4BL, even for the consistent results.
Although we excluded two bug localization techniques in Bench4BL as shown in Section~\ref{sss:selection}, we have not verified the correctness of the remaining four techniques~(and the two derived from them).
The possibility of incorrectness in implementing these bug localization techniques still exists.
However, we believe that these four implementations have a certain degree of correctness.
As shown in Fig.~\ref{f:bl_technique}, there is a relationship between the techniques in terms of the use of additional information, and it is expected that the more additional information is used, the higher the accuracy becomes.
The accuracy attained in this study did not contradict this relationship, suggesting that the implementations may yield the expected accuracy.

Moreover, it should be noted that Bench4BL can only run on file-level bug localization.
Wang et al.~\cite{wang-issta2015} suggested that the results of most bug localization techniques at the file level still leave developers with a large amount of code to examine.
Therefore, it might be beneficial to conduct the same experiment at the class or method level on different datasets.
Noteworthy is that, although our previous work was conducted on method-level modules, a large-scale method-level benchmark dataset is not yet available.
We continue the discussion on method-level bug localization in Section~\ref{ss:meth-sabl}.

Finally, the accuracy of smell detection in inFusion may be a threat.
Manually validating the smells detected in inFusion to exclude the presence of false positives and negatives remains a future task.
We continue the discussion on the false positives in Section~\ref{ss:accuracy-smell}.

\subsection{External Validity}

Although we mitigated the threat of external validity by using a sufficient number of bug reports, we limited our attention to Java systems.
Moreover, we only used open-source systems.
Therefore, performing similar studies on industrial systems may be beneficial.
In addition, although we used the largest available bug localization dataset in this study, the optimal configuration presented in this paper might have different results on other datasets.
Moreover, we considered 16 types of code smells in this study, yet other types of code smells or other methods to calculate the smell severity~\cite{taba-icsm2013} that were not considered in this study are also available.
In addition, we only used inFusion as the smell detector, despite the existence of other possible smell detectors that were not considered in this study.
As a result, they may have a different effect on bug localization.
Finally, because our experiments on bug localization were only conducted at the file level owing to the limitations of Bench4BL, we would need to conduct experiments at other levels, e.g., the method level.

\subsubsection{Conclusion Validity}

Although we performed statistical tests~(Wilcoxon signed-rank tests) and confirmed statistical significance between bug localization techniques, the obtained effect size computed via Cliff's delta was very small.
This indicates a possibility that the significance came from a large sample set in our experiment, and the essential effect might be negligible.
We will discuss this point in Section~\ref{s:accuracy}.


\section{Discussion}\label{s:discussion}

\subsection{Accuracy of Bug Localization Techniques}\label{s:accuracy}

Even though our results improved significantly by using information about the code smells in combination with the bug localization technique, one might consider the improvements to be small, i.e., improvements of 10.4--23.0\% in relative comparison.
Actually, the obtained effect size computed via Cliff's delta was all little.
However, we argue that these improvements are meaningful in the context of bug localization.
As shown in Fig.~\ref{f:best-results}, a comparison of the accuracy of each baseline technique reveals that the improvements are on the same scale.
For example, even though \BLT{BugLocator} was improved relative to \BLT{rVSM}, MAP improved by only 6.5\% in relative comparison~(from 0.391 to 0.417; 0.025 in absolute comparison).
This means that improving the bug localization accuracy is generally difficult, even if additional information sources, such as past bug reports or history information, were to be utilized.
Smell-aware bug localization can significantly improve the accuracy of state-of-the-art bug localization techniques in similar amounts.
In addition, it is noteworthy that the smell-aware bug localization technique uses only the source code and does not require additional information.
Therefore, we suggest that source code characteristics, such as code smells, should be considered when performing bug localization.

\subsection{Accuracy of Detected Smells}\label{ss:accuracy-smell}

In some cases, smell detectors might produce false positives. 
For example, a parser class having \GodClass is not considered problematic because its scope is generally large, and refactoring it might even reduce the comprehensibility of the class~\cite{fontana-saner2016,fontana-catalog2015}.
Several existing studies have reported the accuracy of smells detected by inFusion~\cite{fernandes-ease2016,paiva-jserd2017,fontana-jot2012}, but the reported accuracy varies widely and covers only a small subset of smell types.

To mitigate this threat, the authors manually verified sampled smell instances detected by inFusion.
We followed the false positive catalog by Fontana et al.~\cite{fontana-saner2016,fontana-catalog2015}.
For the seven smell types listed in the catalog with their false positive detection strategy~(\BlobClass, \DataClass, \GodClass, \BlobOperation, \FeatureEnvy, \MessageChains, and \ShotgunSurgery), we randomly selected five smell instances for each type, and we collected 35 smell instances in total.
We selected one instance from each of the top five projects with the highest number of smell instances in total, resulting in five instances for each type.
For each instance, two of the authors independently judged whether it met the condition to be regarded as a false positive according to the false positive detection strategy in the catalog.
In the case of two authors' decisions being in conflict, we conducted a discussion to reach a consensus.
As a result, four false positives out of 35 were identified, yielding a precision of 0.89.
Although the number of extracted samples was very small, this result suggests that a certain percentage of smells used in our study were correct instances to be regarded as smells.
We conclude that false positives may have little effect on the results of this study.

However, because the process of identifying false positives was performed by the authors, who are not the main developers of the projects used for this study, we cannot ensure the completeness of the identified results.
Furthermore, our sampling approach cannot confirm the recall of the detected smells.
Both are still regarded as a threat to validity.%

\subsection{Distribution of $\alpha$ Values}\label{s:alpha-dist}

\begin{figure}[tb]\centering
  \includegraphics[width=0.8\linewidth]{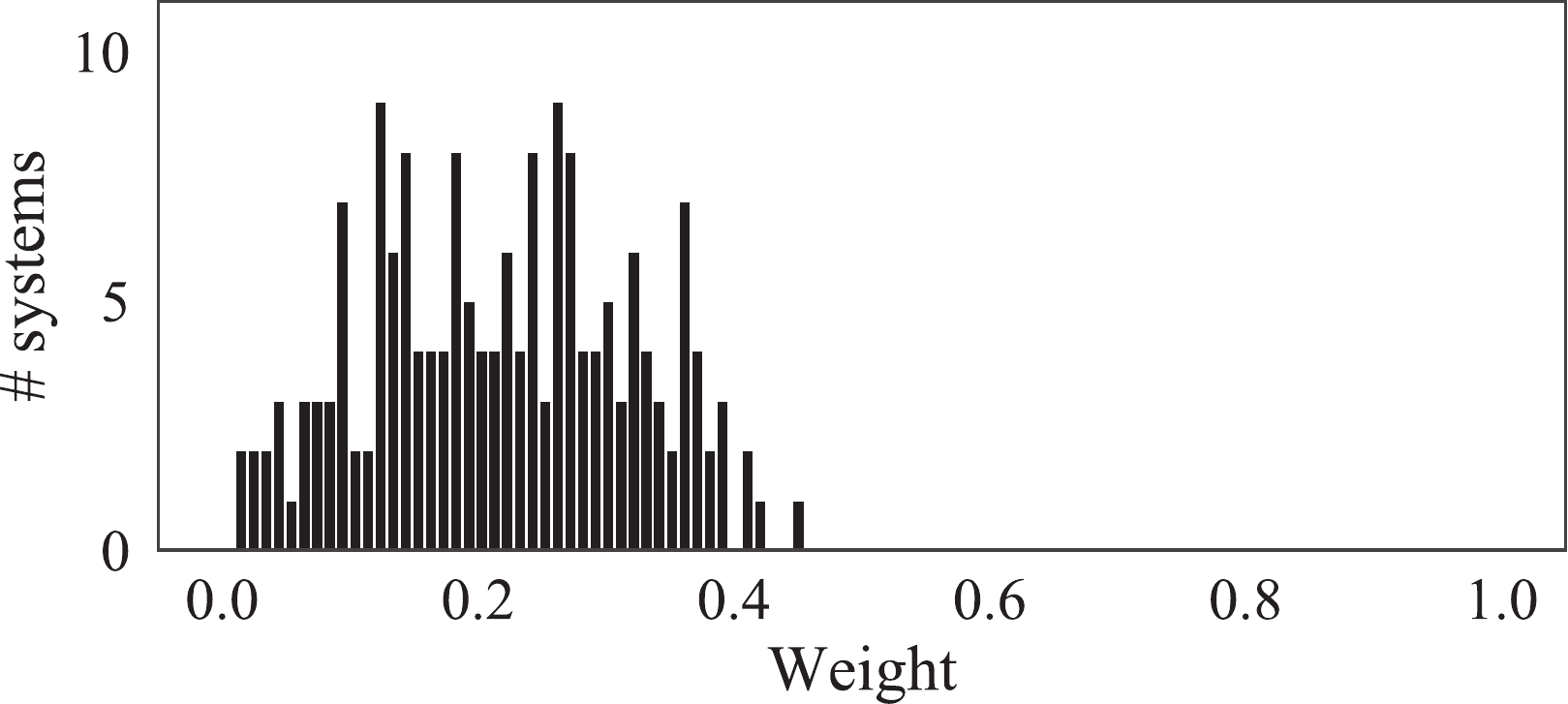}
  \caption{Distribution of the optimal $\alpha$ values.}\label{f:alpha-dist}
\end{figure}
We studied the distribution of the optimal value of $\alpha$ for each system.
We investigated 309 systems using MAP as the evaluation metric when the optimal configuration obtained from \RQ{3}, i.e., \rnum{1}~\Config{\CP{$g_3$}{both class and method levels}, \CP{$a_3$}{existence of smells}, \CP{$s_5$}{well selected smell types}}, was used.
Note that we observed multiple $\alpha$ values that can maximize the MAP value in 85 out of 309 systems, such as \textsf{DATAMONGO 1.1.2} as shown in Fig.~\ref{f:system-plots}.
For the sake of simplicity of the analysis, we excluded these systems and selected the other 224 systems as the target of the subsequent investigation.

The systems were broadly divided into two categories: those who could utilize smell-based scores effectively and those for which the detected smells did not work at all.
On the one hand, in 49 of 224 systems, the setting of $\alpha = 0$ produced their optimal ranking.
This means that any blending of smell-based scores reduced the accuracy of the resulting rankings; i.e., the use of smells did not improve the accuracy of these systems.
On the other hand, in the remaining 175 systems, blending the smell-based score improved the ranking compared to the base IR-based bug localization using VSM.
Figure~\ref{f:alpha-dist} shows a histogram representing the distribution of optimal $\alpha$ values of the 175 systems.
The average value of the obtained optimal $\alpha$ values was 0.215, and all the values were less than 0.5.
For these systems to which the smell information contributed positively, a weak blending of the smell-based score with the IR-based score tends to improve the ranking well.
Although this analysis is limited because of the selected instances of $\alpha$ values to be used, we think that this average value can be regarded as a representative.
The prediction of an appropriate $\alpha$ value, for example, using a machine learning technique, contributes to bug localization improvement.
For example, it may be possible to compute an appropriate $\alpha$ value for a project version by using the results obtained from past versions of the project.
Note that such an approach should deal with the size variation of the versions and some inappropriate versions that smell-aware bug localization is not suitable, as we studied in this section.

\subsection{Application to Method-Level Bug Localization}\label{ss:meth-sabl}

\begin{figure}[tb]\centering
  \newcommand{\figheight}{3cm}
  {\scriptsize\tabcolsep=0.2em\begin{tabular}{cc}
    \includegraphics[height=\figheight]{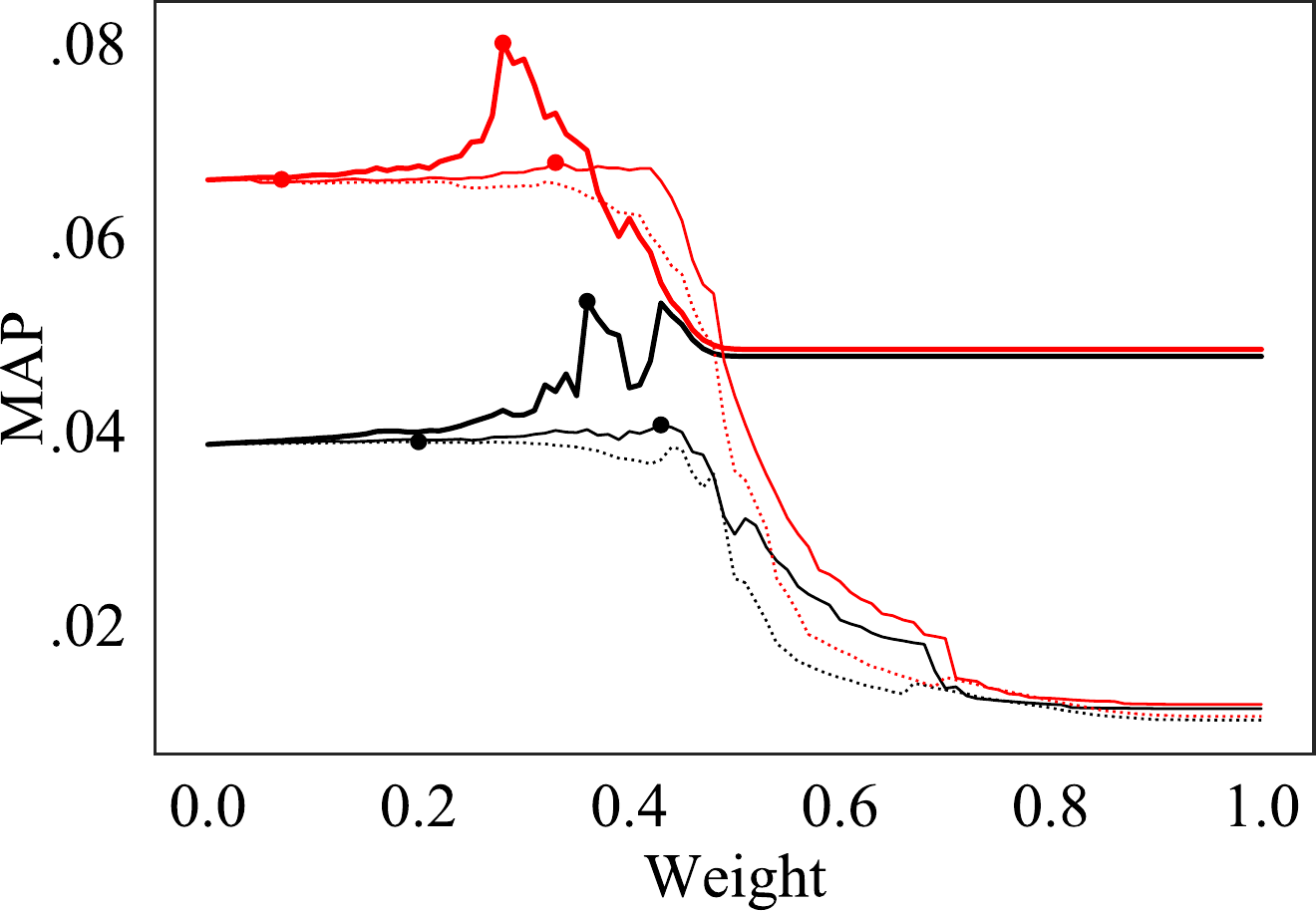} &
    \includegraphics[height=\figheight]{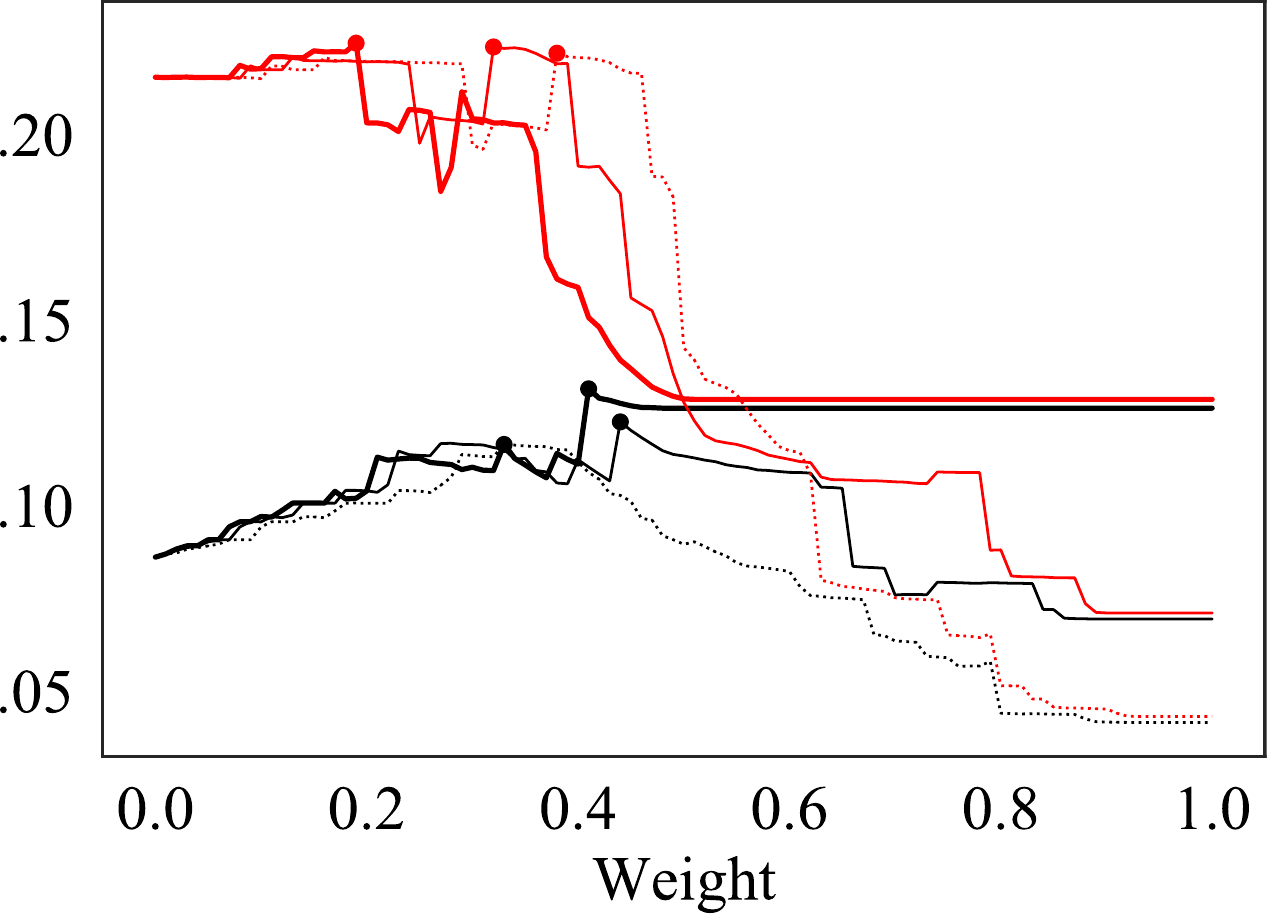}  \\
    (a) \textsf{ArgoUML 0.20}. &
    (b) \textsf{JabRef 2.0}. \\ ~ \\
    \includegraphics[height=\figheight]{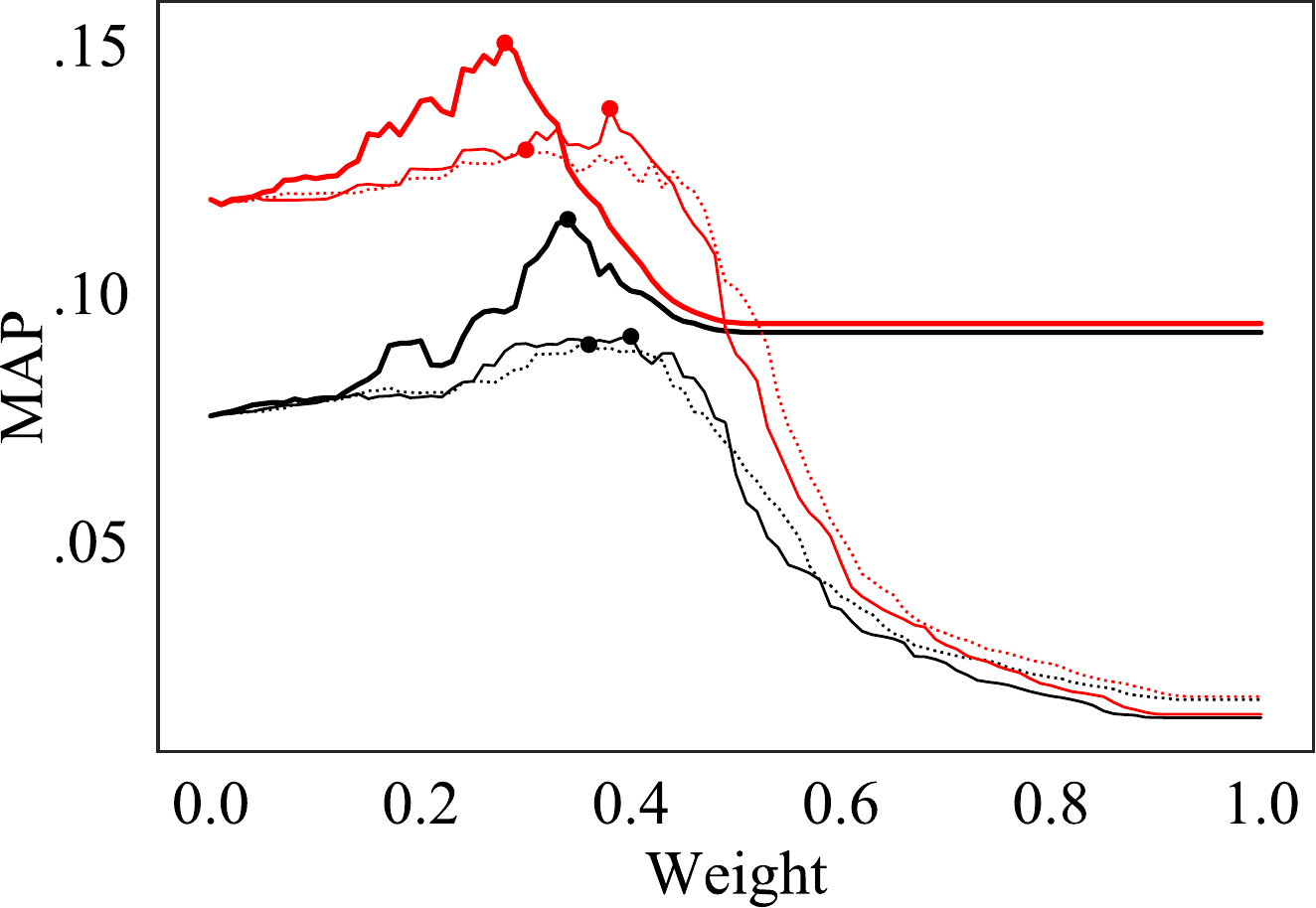} &
    \includegraphics[height=\figheight]{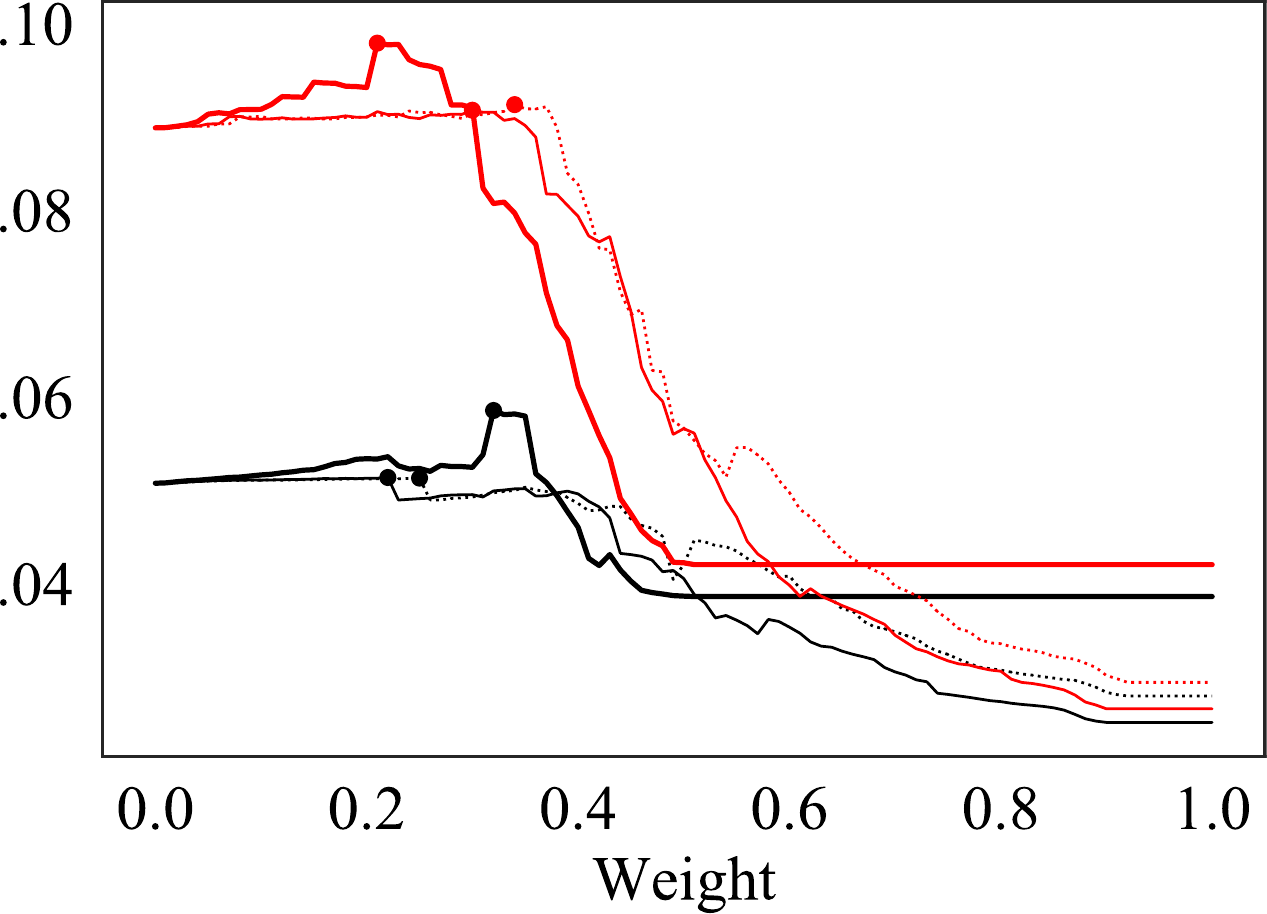} \\
    (c) \textsf{jEdit 4.2}. &
    (d) \textsf{muCommander 0.8}. \\ ~ \\
    \multicolumn{2}{c}{\includegraphics[width=7cm]{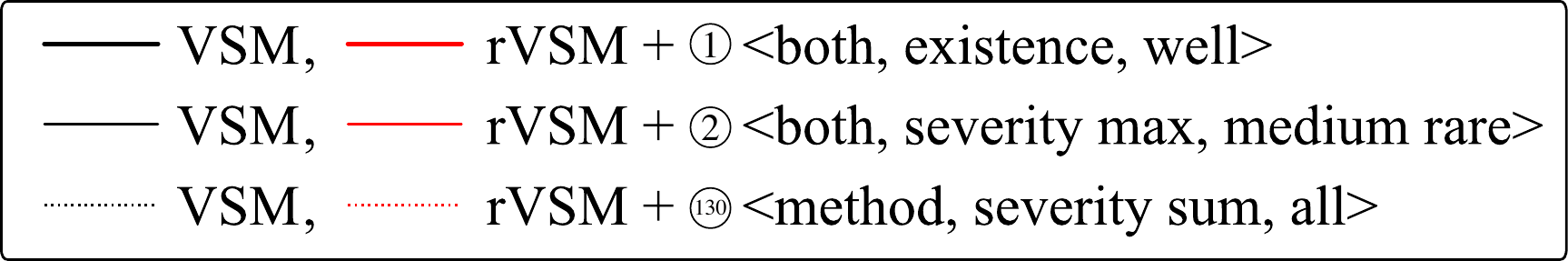}}
  \end{tabular}}
  \caption{Distribution of MAP values in method-level bug localization.}\label{f:meth-plots}
\end{figure}
In our previous paper~\cite{takahashi-icpc2018}, we have studied not only class-level but also method-level bug localization.
However, the study in this paper focused on the class level.
This is because Bench4BL is based on files, which does not provide bug localization at the method level, and no bug localization benchmarking framework supports the application of bug localization techniques at the method level.

To preliminary confirm whether the best configurations obtained in the study are effective for method-level bug localization, we manually adapted four systems used in the previous paper to the Bench4BL framework and obtained method-level application results.
We obtained the method level matching results by converting source code snapshots to those at the method level using FinerGit~\cite{higo-jss2020}.
The results are shown in Fig.~\ref{f:meth-plots}.
Similarly to Fig.~\ref{f:system-plots}, the plots in the figure show the results using the two best configurations~(\rnum{1} and \rnum{2}) and the original configuration at the method level~(\rnum{130}) using \BLT{VSM} and \BLT{rVSM} as the baseline IR techniques.
As we can see from the plots, the use of the best configurations improved the MAP for all systems in both \BLT{VSM} and \BLT{rVSM}.
Although a detailed study on the method level bug localization is subject to future work, these results suggest the applicability of the proposed smell-aware bug localization technique to the method level.


\section{Conclusion}\label{s:conclusion}

In this study, we replicated the work conducted with our previous smell-aware bug localization technique on a large-scale dataset and confirmed significant performance improvement.
We proposed a generalized smell-aware bug localization technique to derive the optimal configurations for code smell information.
We found that the optimal configuration entails the use of granularity that reflects both class- and method-level smells and the maximum severity when aggregating a certainly selected types of code smells or the existence of very limited types of smells.
Finally, we combined our proposed technique with different baseline techniques and found that the performance improved significantly.
These results suggest that code smells can be used to effectively improve existing bug localization techniques without the need for additional information.

Code smell detection does not require more inputs than bug localization in general.
Although the improvement was slight, the application of the smell-aware approach can improve bug localization, which is applicable in many situations.
Our study also revealed that there are situations where the effect of the smell-aware approach was negative.
It is desirable to develop further techniques that use smell information more effectively to avoid losing the accuracy of the baseline bug localization technique.
The use of machine learning techniques or other data fusion techniques might be effective rather than a simple linear combination that we used in this paper.

In the future, we also aim to specify the value of $\alpha$ for each version of a specific project.
For example, we can set the value of $\alpha$ based on the optimal value of previous versions or by using a machine learning approach.

An appendix including the experimental materials is available in online~\cite{appendix}.


\section*{Acknowledgments}
This work was partly supported by JSPS KAKENHI Grant Number JP18K11238.


\end{document}